\documentclass[useAMS,usegraphicx,usenatbib]{mn2e} 
\usepackage{longtable}
\usepackage{natbib}
\usepackage[colorlinks=True, urlcolor=blue, citecolor=blue]{hyperref}

\usepackage{graphicx}
\usepackage{graphics}
\usepackage{float}
\usepackage{booktabs}
\usepackage{caption}
\usepackage{subcaption}
\usepackage{csquotes}
\usepackage{amsmath}
\usepackage{color}
\usepackage{url}
\usepackage{ulem}
\usepackage{multirow}
\usepackage{amsmath}
\usepackage{wasysym}
\usepackage{amssymb}
\usepackage{pifont}

\bibpunct{(}{)}{;}{a}{}{,}

\def\apjl{ApJ}%
\def\apjs{ApJS}%
\def\ao{Appl.~Opt.}%
\def\aap{A\&A}%
\title[BASS XXVII: Scattered X-ray radiation in obscured AGN]{BAT AGN Spectroscopic Survey XXVII: Scattered X-Ray Radiation in Obscured Active Galactic Nuclei}
\author[Gupta et al.]{K. K. Gupta$^{1}$\thanks{E-mail:
kriti.gupta@mail.udp.cl}, C. Ricci$^{1,2,3}$, A. Tortosa$^{1}$, Y. Ueda$^{4}$, T. Kawamuro$^{1,5}$, M. Koss$^{6}$,
\newauthor B. Trakhtenbrot$^{7}$, K. Oh$^{4,8}$, F. E. Bauer$^{9,10,11}$, F. Ricci$^{9}$, G. C. Privon$^{12,13}$, 
\newauthor L. Zappacosta$^{14}$, D. Stern$^{15}$, D. Kakkad$^{16}$, E. Piconcelli$^{14}$, S. Veilleux$^{17}$, 
\newauthor R. Mushotzky$^{17}$, 
T. Caglar$^{18}$, K. Ichikawa$^{19}$, A. Elagali$^{20,21}$, M. C. Powell$^{22}$, 
\newauthor C. M. Urry$^{23}$, and F. Harrison$^{24}$\\
Affiliations can be found after the references.
}
\begin{document}

\date{Received; accepted}

\pagerange{\pageref{firstpage}--\pageref{lastpage}} \pubyear{2021}

\maketitle

\label{firstpage}

\begin{abstract}
Accreting supermassive black holes (SMBHs), also known as active galactic nuclei (AGN), are generally surrounded by large amounts of gas and dust. This surrounding material reprocesses the primary X-ray emission produced close to the SMBH and gives rise to several components in the broadband X-ray spectra of AGN, including a power-law possibly associated with Thomson-scattered radiation. In this work, we study the properties of this scattered component for a sample of 386 hard-X-ray-selected, nearby ($z \sim 0.03$) obscured AGN from the 70-month \textit{Swift}/BAT catalog. We investigate how the fraction of Thomson-scattered radiation correlates with different physical properties of AGN, such as line-of-sight column density, X-ray luminosity, black hole mass, and Eddington ratio. We find a significant negative correlation between the scattering fraction and the column density. Based on a large number of spectral simulations, we exclude the possibility that this anti-correlation is due to degeneracies between the parameters. The negative correlation also persists when considering different ranges of luminosity, black hole mass, and Eddington ratio. We discuss how this correlation might be either due to the angle dependence of the Thomson cross-section or to more obscured sources having a higher covering factor of the torus. We also find a positive correlation between the scattering fraction and the ratio of [O{\scriptsize\,III}] $\lambda 5007$ to X-ray luminosity. This result is consistent with previous studies and suggests that the Thomson-scattered component is associated with the narrow-line region.

\end{abstract}	
               



\section{Introduction}

Supermassive black holes (SMBHs), present in the centers of virtually all massive galaxies (e.g., \citealp{doi:10.1146/annurev.aa.33.090195.003053}), during their accreting phase are known as active galactic nuclei (AGN).  The growth of SMBHs and that of their host galaxies is thought to be closely connected since several of their properties are strongly correlated (e.g., \citealp{1998AJ....115.2285M}; \citealp{2003ApJ...589L..21M}). These include the tight correlations found between the stellar velocity dispersion of the galaxy and the mass of the central black hole (e.g., \citealp{2000ApJ...539L...9F}; \citealp{2000ApJ...539L..13G}; \citealp{2002ApJ...574..740T}; \citealp{2020A&A...634A.114C}). Hence, AGN are believed to play an important role in the evolution of their host galaxies (e.g., \citealp{doi:10.1146/annurev-astro-082708-101811}).  

AGN are usually surrounded by a large amount of gas and dust, distributed in a structure referred to as the torus (e.g., \citealp{1995PASP..107..803U}). This torus is thought to be responsible for reprocessing a fraction of the light emitted by the accretion flow (e.g., \citealp{1993ARA&A..31..473A}). The accretion disk, which is the main source of radiation in non-jetted AGN, emits optical/UV photons, which in some sources can be completely obscured by the surrounding torus. In fact, previous studies have shown that a majority of AGN are obscured (e.g., \citealp{1998A&A...338..781M}; \citealp{1999ApJ...522..157R}; \citealp{Ricci_2015}; \citealp{2018ApJ...854...49M}) and are thus difficult to account for based on optical or UV surveys. These obscured AGN are believed to contribute significantly to the cosmic X-ray background (CXB; e.g., \citealp{1995A&A...296....1C}; \citealp{2003ApJ...598..886U}; \citealp{2007A&A...463...79G}; \citealp{2014ApJ...786..104U}; \citealp{2020ApJ...889...17A}). Therefore, the study of obscured SMBHs is essential to improve our understanding of the entire AGN population and their evolution. This can be done by making use of a characteristic feature of AGN, which is their strong X-ray emission, which is produced by the Comptonization of optical and UV photons in a corona of hot electrons located close to the SMBH (e.g., \citealp{1991ApJ...380L..51H}). This energetic radiation, and in particular that emitted in the hard-X-ray band ($E >$ 10 keV), can be used to find obscured AGN, and to study their properties, because: (a) they are significantly less absorption-biased than lower-energy X-rays, hence can penetrate large column densities ($N_{\rm H}$), up to Compton-thick values ($N_{\rm H}>10^{24}\,{\rm cm}^{-2}$; see Figure 1 of \citealp{Ricci_2015}) and (b) they are substantially less contaminated by the light associated with the host galaxy than softer X-rays.

The broadband X-ray spectra of AGN have already been analysed in great detail in many previous studies (e.g., \citealp{2007A&A...461.1209D}; \citealp{2009ApJ...690.1322W};
\citealp{2016ApJ...825...85K}; \citealp{2017ApJS..233...17R}). Typically, in unobscured AGN, the spectrum is dominated by a primary X-ray continuum associated with radiation produced by the corona (e.g., \citealp{2008A&A...485..417D}; \citealp{2009A&A...505..417B}). Apart from this, the spectrum also shows several distinctive features that can be attributed to the absorption and reprocessing of the primary X-ray radiation by the circumnuclear material. For example, for column densities above $10^{22}\,{\rm cm}^{-2}$, the primary X-ray continuum is significantly suppressed due to photoelectric absorption. Signatures of reprocessed X-ray radiation by the circumnuclear material are also seen in the X-ray spectrum in the form of a broad \enquote{Compton hump} peaking around 30 keV (e.g., \citealp{1988ApJ...335...57L}; \citealp{1994MNRAS.267..743G}; \citealp{1994ApJ...420L..57K}) and a fluorescent Fe K$\alpha$ line at 6.4 keV (e.g., \citealp{2000PASP..112.1145F}; \citealp{2004ApJ...604...63Y}; \citealp{2010ApJS..187..581S}; \citealp{2011ApJ...727...19F}; \citealp{2014MNRAS.441.3622R}). These features provide important information about the geometry, as well as the physical properties of the circumnuclear material (e.g., \citealp{1991MNRAS.249..352G}; \citealp{1991A&A...247...25M}; \citealp{2013A&A...549A..72P}). 

When the X-ray continuum is obscured ($N_{\rm H} \gtrsim 10^{22}\,{\rm cm}^{-2}$), a component possibly related to Thomson scattering of the X-ray continuum by circumnuclear photoionized gas also becomes visible. This component is believed to be produced in the Compton-thin circumnuclear material and is observed as a weak, unabsorbed, power-law continuum emerging at energies lower than the photoelectric cutoff. The scattered X-ray radiation can help us to understand the effects of obscuration and can shed light on the structure of the inner regions of AGN. The parameter generally used to study the scattered radiation in AGN is the \enquote{scattering fraction} ($f_{\rm scatt}$; e.g., \citealp{1997ApJS..113...23T}; \citealp{2006A&A...446..459C}; \citealp{2007ApJ...664L..79U}; \citealp{2009ApJ...690.1322W}; \citealp{2020ApJ...897..107Y}). This parameter denotes the fraction of radiation scattered with respect to the primary X-ray emission of the AGN. Its intensity depends on both the covering angle of the surrounding torus and the amount of gas available for scattering (e.g., \citealp{2002ApJ...573L..81L}; \citealp{2009ApJ...696.1657E}).

Using broadband X-ray observations of hard-X-ray-selected AGN, \cite{2007ApJ...664L..79U} discovered a new type of AGN with extremely low values of scattering fraction ($f_{\rm scatt} < 0.5$\%). They concluded that either the central SMBH in these sources is buried in a geometrically-thick torus with a small opening angle, or the amount of gas available for scattering is exceptionally low, or both. \cite{2009ApJ...705..454N} later presented an improved sample of these \enquote*{buried} AGN, and investigated their multi-wavelength properties \citep{2010ApJ...711..144N}, finding that these objects also tend to have lower values of the ratio of [O\,{\scriptsize III}] $\lambda$5007 to X-ray luminosity ($L_{\rm [O\,III]}/L_{\rm X}$). They interpreted this as evidence of these AGN having a small opening angle of the torus that collimates the narrow-line region (NLR), a result further supported by \cite{2015ApJ...815....1U} and \cite{2016ApJS..225...14K}. Studies have already shown that the soft X-ray ($E <$ 10 keV) emission in obscured sources can extend up to hundreds of parsecs (e.g., \citealp{2000ApJ...543L.115S}; \citealp{2001ApJ...556....6Y}; \citealp{2014ApJ...788...54G}) and has a morphology similar to that of the optical NLR, as traced by the [O\,{\scriptsize III}] $\lambda$5007 emission (e.g., \citealp{2006A&A...448..499B}; \citealp{2010A&A...516A...9D}; \citealp{2010MNRAS.405..553B}; \citealp{2017MNRAS.469.2720G}; \citealp{2018ApJ...865...83F}). The detailed analysis of high-resolution X-ray spectra of obscured AGN (e.g., NGC 1068 by \citealp{2002ApJ...575..732K} and \citealp{2002A&A...396..761B}; Mrk 3 by \citealp{2000ApJ...543L.115S}) showed that the $<$ 3--4 keV range is dominated by photoionization lines. This points towards both the soft X-ray and the optical emission lines being produced in the photoionized gas of the NLR. As a result, low values of $L_{\rm [O\,III]}/L_{\rm X}$ for buried AGN suggest that the possible locus of Thomson scattering of X-ray radiation is also in the NLR. However, \cite{2012ApJ...754...45I}, who presented a mid- to far-infrared study of a sample of local AGN, attributed the low scattering values of AGN to strong starburst activities of the host galaxy. On the other hand, \cite{2014MNRAS.438..647H} found that objects with low scattering fractions reside in high-inclination galaxies or merger systems. Therefore, they concluded that the low values of $f_{\rm scatt}$ are related to the obscuration of scattered emission due to host galactic gas and dust. 

This work aims to study the properties of Thomson-scattered radiation in a sample of hard-X-ray-selected AGN and to constrain the possible region of origin of this radiation. The objects in our sample were detected by the Burst Alert Telescope (BAT: \citealp{2005SSRv..120..143B}; \citealp{2013ApJS..209...14K}) on board the NASA mission, \textit{Neil Gehrels Swift Observatory}. BAT operates in the 14--195 keV energy band and is therefore sensitive to heavily obscured AGN. Since 2005, BAT has been continuously surveying the entire sky and has proved very useful to study AGN in the local Universe ($z$ $<$ 0.1). The BAT AGN Spectroscopic Survey (BASS\footnote{\textcolor{blue}{\url{www.bass-survey.com}}}) has provided high-quality multi-wavelength data for the BAT AGN, including black hole mass measurements \citep{2017ApJ...850...74K}, X-ray spectroscopy and modelling \citep{2017ApJS..233...17R}, near-infrared (NIR; 1--2.4$\,\micron$) spectra and modeling \citep{2017MNRAS.467..540L}, NIR AO imaging \citep{2018Natur.563..214K}, extensive continuum modeling of the far-infrared emission \citep{2019ApJ...870...31I}, radio emission \citep{2020MNRAS.492.4216S}, and molecular gas measurements \citep{2020arXiv201015849K}. From this data, numerous correlations have been found such as between X-ray emission and high ionization optical lines (e.g., \citealp{2015MNRAS.454.3622B}) and outflows (e.g., \citealp{2020MNRAS.491.5867R}) with the Eddington ratio being a key parameter in these trends (e.g. \citealp{2017Natur.549..488R}; \citealp{2017MNRAS.464.1466O}; \citealp{2018MNRAS.480.1819R}).

We use the 70-month \textit{Swift}/BAT catalog \citep{2013ApJS..207...19B}, with revised counterpart classifications from \cite{2017ApJS..233...17R}, to understand the properties of scattered X-ray radiation in nearby obscured AGN and how it relates with the physical properties of AGN. To do so, we study possible correlations between scattering fraction and other physical properties of accreting SMBHs, such as their column density ($N_{\rm H}$), X-ray luminosity ($L_{\rm X}$), black hole mass ($M_{\rm BH}$), and Eddington ratio ($\lambda_{\rm Edd}$). We also investigate the dependence of $f_{\rm scatt}$ on the ratio of [O\,{\scriptsize III}] $\lambda$5007 luminosity to X-ray luminosity ($L_{\rm [O\,III]}/L_{\rm X}$), as well as the ratio of [O\,{\scriptsize II}] $\lambda$3727 luminosity to X-ray luminosity ($L_{\rm [O\,II]}/L_{\rm X}$).

The structure of the paper is as follows. In Section \ref{sect:sampledata} we describe the sample used in this work, while in Section \ref{sect:corr1} we present the correlation we found between scattering fraction and column density. In Section \ref{sect:deg}, we simulate and fit spectra for a dummy population of obscured AGN, to verify that the correlation found in this section is intrinsic to our sample, and not due to parameter degeneracy. The correlations with $L_{\rm [O\,III]}/L_{\rm X}$ and $L_{\rm [O\,II]}/L_{\rm X}$ are discussed in Section \ref{sect:corr2}, while potential correlations with black hole mass, X-ray luminosity, and Eddington ratio are investigated in Section \ref{sect:corr3}. Finally, we discuss and summarize our findings in Sections \ref{sect:discussion} and \ref{sect:summary}. Throughout the paper, we assume a cosmological model with $H_{\rm 0}=70\,\rm km\,s^{-1}\,Mpc^{-1}$, $\Omega_{\rm M}=0.3$, and $\Omega_{\Lambda}=0.7$.


\section{Sample and Data}\label{sect:sampledata}

We use here the 70-month \textit{Swift}/BAT catalog, which includes 838 hard-X-ray-selected, nearby (median redshift $=0.037$) AGN. Of these, 386 ($\sim$53\%) sources are classified as obscured AGN based on strong intrinsic absorption required in the initial absorbed power-law fits to their X-ray spectra (\citealp{2017ApJS..233...17R}; see Section \ref{sect:xraydata}). The line-of-sight column densities ($N_{\rm H}$) of these sources lie in the range $10^{21.5}\,{\rm cm}^{-2}$ to $10^{25}\,{\rm cm}^{-2}$. Our analysis focuses on these obscured sources, which show features of Thomson-scattered radiation in their X-ray spectra. Of these, nine sources are flagged as blazars by \citeauthor{2017ApJS..233...17R}(\citeyear{2017ApJS..233...17R}; see also \citealp{2019ApJ...881..154P}). However, they are included in the analysis as their X-ray spectra showed signatures of reprocessed X-ray emission \citep{2017ApJS..233...17R}. The various parameters used in our work have been obtained from the optical (Section \ref{sect:opticaldata}) and X-ray spectroscopic analysis (Section {\ref{sect:xraydata}}) of these sources, carried out as part of the first data release (DR1) of BASS (e.g., \citealp{2017ApJ...850...74K,2017ApJS..233...17R}).


\subsection{X-Ray Data} \label{sect:xraydata}

The \textit{Swift}/BAT AGN have been extensively followed up by X-ray observatories covering the 0.3--10 keV energy range. By combining observations from \textit{XMM-Newton} {\citep{2001A&A...365L...1J}}, \textit{Swift}/XRT \citep{2005SSRv..120..165B}, \textit{ASCA} \citep{1994PASJ...46L..37T}, \textit{Chandra} \citep{2000SPIE.4012....2W} and \textit{Suzaku} \citep{2007PASJ...59S...1M} in the soft X-ray band ($E < 10$ keV), together with the \textit{Swift}/BAT data, \citeauthor{2017ApJS..233...17R} (\citeyear{2017ApJS..233...17R}; hereafter R17) carried out a detailed broadband X-ray (0.3--150 keV) spectral analysis of all 838 AGN from the 70-month \textit{Swift}/BAT catalog, using {\small XSPEC} (\citealp{1996ASPC..101...17A}; see Section \ref{sect:deg}). Firstly, they fitted the X-ray spectra of all sources using a simple power-law model, including Galactic absorption along the line of sight (\texttt{TBABS}\textsubscript{\texttt{Gal}}.[\texttt{ZPOW}] in {\small XSPEC}). The residuals from these fits were then visually inspected to check for signatures of absorption from neutral matter to classify the sources as obscured or unobscured. After this preliminary exercise, 24 different spectral models were used for the X-ray spectral fitting. 

The sample of 386 obscured, non-blazar AGN were fitted using nine models (B1 to B9; see R17 for more details) of increasing complexity. The different components included in these models are: (1) an X-ray continuum with a high energy cutoff (\texttt{CUTOFFPL}) representing the primary X-ray emission from the corona, (2) absorption of the X-ray radiation by neutral material via photoelectric absorption (\texttt{ZPHABS}) and Compton scattering (\texttt{CABS}), (3) reflection of the primary X-ray continuum by optically thick, neutral circumnuclear material (\texttt{PEXRAV}; \citealp{1995MNRAS.273..837M}), and (4) Thomson scattering of part of the primary X-ray radiation by Compton-thin circumnuclear material. 

For 272/386 (70\%) obscured sources, model B1 was sufficient and hence adopted by R17 for the spectral fitting. This model consisted of an absorbed primary X-ray continuum, along with an unobscured reflection component and a Thomson-scattered component. For the latter, a cutoff power-law was used, with values of the free parameters (photon index [$\Gamma$], cutoff energy [$E_{\rm C}$], normalization [$K$]) set to be the same as those of the primary X-ray continuum. A multiplicative constant corresponding to the scattering fraction ($f_{\rm scatt}$) was added to this component, as a free parameter, to renormalize the flux. For the remaining objects, model B1 was modified by adding multiple collisionally ionized plasma components or partially covering absorbers, as and when such a modified model improved the fit significantly.

The best-fit values of all free parameters in different models, employed for the spectral fitting of the AGN sources, are reported by R17. They also provide absorption-corrected fluxes of the primary continuum emission in the 2--10 keV, 14--150 keV, and 14--195 keV energy bands ($F_{\rm 2-10}$, $F_{\rm 14-150}$, and $F_{14-195}$, respectively) for all obscured sources. The main spectral parameters used in our analysis, such as $f_{\rm scatt}$, $N_{\rm H}$, and $F_{\rm X}$, have all been obtained from their fitting analysis. Best-fit values of column densities, along with upper and lower bounds (of 90\% confidence level), are available for all 386 obscured sources. In the case of the scattering fraction, best-fit values along with upper and lower errors (at 90\% confidence level) are available for 250/386 ($\sim$65\%) objects, while for the rest (136/386 $\sim$ 35\%), upper limits are provided. We also calculate the intrinsic (i.e., absorption-corrected and k-corrected) luminosities for the obscured sources in the 2--10 keV, 14--150 keV, and 14--195 keV energy bands ($L_{\rm 2-10}$, $L_{\rm 14-150}$, and $L_{14-195}$, respectively). Spectroscopic redshifts and corresponding distance values are available for 382/386 ($\sim$99\%) sources, and therefore, we can calculate the intrinsic luminosity for these 382 sources.


\subsection{Optical Data} \label{sect:opticaldata}

\citeauthor{2017ApJ...850...74K} (\citeyear{2017ApJ...850...74K}; BASS DR1) and Koss et al. (in prep.; BASS DR2) analysed the optical spectra of the 838 hard-X-ray-selected AGN in the 70-month \textit{Swift}/BAT catalog. They performed multiple spectral measurements to understand the general properties of \textit{Swift}/BAT AGN, such as broad and narrow emission-line diagnostics to classify these AGN. They also fitted the host galaxies using stellar templates to calculate their stellar velocity dispersions. They presented the fluxes and strengths of optical emission lines, as well as black hole mass estimates and accretion rates, for both obscured and unobscured AGN in their sample. Out of the 386 sources in our sample, reliable black hole mass estimates that we adopt in this work are available for 273 ($\sim$71\%) sources. For 261/273 ($\sim$96\%) sources, the black hole masses have been derived from the correlation between the black hole mass and the stellar velocity dispersion of the host galaxy given by \cite{2013ARA&A..51..511K}. For the remaining (12/273 $\sim$ 4\%) sources, we use black hole mass values from the literature. We calculate the Eddington ratio ($\lambda_{\rm Edd}$) for these 273 sources by incorporating their black hole masses and the 2--10 keV bolometric correction ($L_{\rm bol} = 20 \times L_{2-10}$) following \cite{2009MNRAS.399.1553V}:

\begin {equation} \label{eq:1}
\lambda_{\rm Edd} = \left[\frac{20 \times L_{2-10}\,(\rm erg\,s^{-1})}{1.5\times10^{38}\times M_{\rm BH}\,(M_{\odot})}\right].
\end{equation}\\
To include the effects of using more advanced bolometric corrections in our work, we repeat part of the analysis described in Section \ref{sect:corr3} with Eddington ratio-dependent bolometric corrections \citep{2009MNRAS.392.1124V} and get similar results.

Finally, in our analysis, we use the emission line measurements presented by \cite{2017ApJ...850...74K} and Oh et al. (in prep.) using the spectral line fitting codes described by \cite{2006MNRAS.366.1151S} and improved by \citeauthor{2011ApJS..195...13O} (\citeyear{2011ApJS..195...13O}, \citeyear{2015ApJS..219....1O}). We adopt a Gaussian amplitude to noise ratio (A/N) of 3 as the threshold for reliable line measurements and hence, do not include line strengths with A/N $< 3$ in our analysis. We use the observed fluxes of the [O\,{\scriptsize II}] $\lambda$3727 line to calculate the observed [O\,{\scriptsize II}] $\lambda$3727 luminosity ($L_{\rm [O\,II]}$) for 295/386 ($\sim$ 76\%) sources. We also use equivalent widths and observed fluxes of the [O\,{\scriptsize III}] $\lambda$5007 line for 359/386 ($\sim$ 93\%) sources and fluxes of the narrow H$\alpha$ and H$\beta$ lines for 318/386 ($\sim$ 82\%) sources and calculate the corresponding observed [O\,{\scriptsize III}] $\lambda$5007 luminosity ($L_{\rm [O\,III]}$) as well as the extinction-corrected [O\,{\scriptsize III}] $\lambda$5007 luminosity ($L_{\rm [O\,III]}^{\rm corr}$) using the Balmer decrement, as described in \cite{2015ApJ...815....1U}.


\begin{figure*}
\begin{center}
\includegraphics[width=1\textwidth]{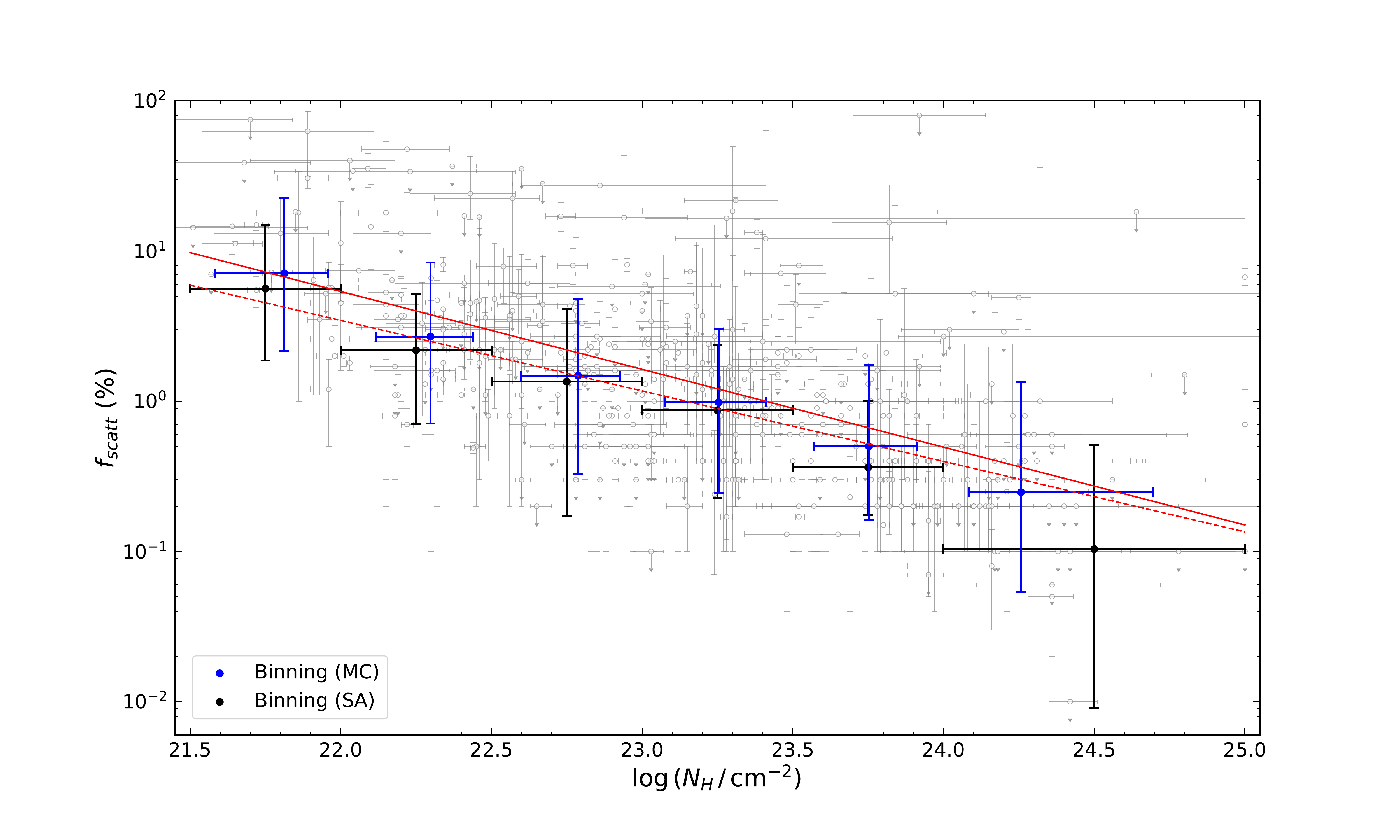}
  \caption{Scattering fraction vs column density: The plot shows the scattering fraction and the column density values (from R17) for our sample of 386 obscured AGN in the background as grey open circles (upper limits as downward arrows and best-fit values with error bars showing errors at 90\% confidence level). Black circles with error bars correspond to the survival analysis (SA) results for each ${\rm log}\,N_{\rm H}$ bin ($f_{\rm scatt}$ median calculated using the 50th percentile and the 1$\sigma$ uncertainty calculated using the 16th and 84th percentile). Blue circles with error bars (1$\sigma$) depict the results obtained from 10,000 Monte Carlo (MC) simulations. The solid red line is the linear regression obtained from the data points excluding upper limits, while the dashed red line is the linear regression calculated using MC simulations. 2 sources in our sample have ${\rm log}(N_{\rm H}/{\rm cm}^{-2}) > 25$, but for display purpose only they are plotted at ${\rm log}(N_{\rm H}/{\rm cm}^{-2}) = 25$. The plot clearly shows a negative correlation between $f_{\rm scatt}$ and ${\rm log}\,N_{\rm H}$. Correlation parameters for the two methods are given in Table \ref{tab:results1}.}
\label{fig:fscatt_vs_logNH}
\end{center}
\end{figure*}


\section{The Correlation Between Scattering Fraction and Column Density}\label{sect:corr1}

We study here the relation between the fraction of Thomson-scattered X-ray radiation and the line-of-sight column density of AGN. In Figure \ref{fig:fscatt_vs_logNH}, we plot the scattering fraction as a function of column density (in grey) for the 386 obscured sources in our sample. The figure clearly shows a negative correlation between these parameters, which we quantify by fitting a linear regression line (in red) to the data points. We initially excluded the upper limits while calculating the regression line here since they can affect our results considerably if not handled properly. However, we do include them in later parts of this analysis to properly quantify their contribution to the correlation. We use the \texttt{linregress}\footnote{\textcolor{blue}{\url{https://docs.scipy.org/doc/scipy/reference/generated/scipy.stats.linregress.html}}} correlation function of the \texttt{scipy.stats} module in python to fit the data with a linear regression. This function minimises the sum of squares of the difference between the actual data points and that predicted by the regression model. Throughout this paper, we use this technique to fit linear regression lines to our data. Slope and intercept of the linear regression line, Pearson's correlation coefficient, and the probability of the data arising from an underlying uncorrelated distribution (null-hypothesis) are reported in Table \ref{tab:results1}. To better visualize the decrease of the scattering fraction as the column density increases, we divide the column density into six bins (of width ${\rm log}[N_{\rm H}/\rm cm^{-2}] = 0.5$, except for the last one) and plot the median of the scattering fraction, along with the standard error (in black), for each bin. The median and error in $f_{\rm scatt}$ in each ${\rm log}\,N_{\rm H}$ bin are calculated using survival analysis (SA; e.g., \citealp{1985ApJ...293..192F};  \citealp{2017MNRAS.466.3161S}) to include the contributions from the upper limits. The median $f_{\rm scatt}$ is calculated using the 50th percentile, and the uncertainties for each bin are calculated with respect to the median value in that bin using the 16th and 84th percentile. For ${\rm log}\,N_{\rm H}$, we simply plot the midpoint of each bin with error bars = 0.5 (= 1 for the last bin).

The SA approach efficiently includes the upper limits in $f_{\rm scatt}$ while calculating its median value in specified bins of column density. However, it does not take into account the errors in $f_{\rm scatt}$ and ${\rm log}\,N_{\rm H}$. In order to include these errors in our analysis, we make use of Monte Carlo (MC) simulations. To apply this technique, we handled the best-fit values with errors and upper limits separately, as discussed below:

\begin{enumerate}
\item Best-Fit Values with Errors: We created an asymmetric Gaussian distribution (normalized) to simulate the values of parameters $f_{\rm scatt}$ and ${\rm log}\,N_{\rm H}$. We used an asymmetric distribution because the upper and lower errors in both the parameters can be unequal. To create this distribution, we first assumed two different symmetric Gaussian distributions with the mean of both Gaussians equal to the best-fit value of the parameter, whereas the standard deviation (1$\sigma$) was chosen to be $\frac{\rm Upper\,\,Error}{1.64}$ and $\frac{\rm Lower\,\,Error}{1.64}$ for the first and second Gaussian, respectively. The errors were divided by 1.64 because they are 2$\sigma$ errors. The final Gaussian was obtained by combining these two Gaussian distributions, such that the probability distribution of values greater than the mean is governed by the first Gaussian, while the probability distribution of values lower than the mean is decided by the second Gaussian. Hence, corresponding to each source with a best-fit $f_{\rm scatt}$ and ${\rm log}\,N_{\rm H}$ value, we created two asymmetric Gaussians depicting the probability distribution functions of $f_{\rm scatt}$ and ${\rm log}\,N_{\rm H}$, by incorporating their best-fit values as well as their upper and lower errors.
\item Best-Fit Upper Limits: Throughout our analysis, upper limits are only present for scattering fraction. In this particular case, we employed a normalized uniform distribution to describe the probability distribution of $f_{\rm scatt}$. The distribution, used for each source with an upper limit, ranges from 0 to the corresponding upper limit value of $f_{\rm scatt}$, with each value in the distribution assigned equal probability. 
\end{enumerate}


After finalizing the probability distribution functions for both parameters ($f_{\rm scatt}$ and ${\rm log}\,N_{\rm H}$), we performed 10,000 runs of MC simulations. In each run, a random value of the scattering fraction and column density was taken for each source from the probability distributions described above. In the case of the column density, values greater than ${\rm log}(N_{\rm H}/{\rm cm}^{-2}) = 26$ were not permitted in the simulations, and for the scattering fraction, negative values were not considered. Thus, we end up with 10,000 simulated values of the scattering fraction and column density for each of the 386 sources. Next, we bin the 3,860,000 simulated column density values into six separate bins (similar to the ${\rm log}\,N_{\rm H}$ bins created for SA) and calculate the median column density for each bin. Values of ${\rm log}(N_{\rm H}/{\rm cm}^{-2}) > 25$ are included in the last bin. Similarly, the corresponding simulated values of $f_{\rm scatt}$ are also binned, and the median value of $f_{\rm scatt}$ in each bin is calculated. The standard error in each bin is calculated using the 16th and 84th percentile, as was done during the SA method. MC results are shown in blue in Figure \ref{fig:fscatt_vs_logNH}. As evident from the figure, median and error values from SA and MC simulations are completely consistent within their error bars. Although, median $f_{\rm scatt}$ values calculated using SA are slightly lower than those calculated from MC simulations. However, unlike SA, MC simulations include the already calculated uncertainties in both parameters (reported by R17) and thus give better estimates of the final median and error in each column density bin.


\begin{table*}
\centering
\caption{Correlation results for Figure \ref{fig:fscatt_vs_logNH} (Section \ref{sect:corr1}), Figure \ref{fig:fscatt_vs_LOIII} (Section \ref{sect:corr2}), and Figure \ref{fig:fscatt_vs_Edd_Ratio} (Section \ref{sect:corr3}).}
\begin{tabular}{ccccccccc}
\hline
\hline
Parameter (in log) & \multicolumn{2}{c}{Slope\,$^{\rm a}$} & \multicolumn{2}{c}{Intercept\,$^{\rm a}$} & \multicolumn{2}{c}{R-Value\,$^{\rm b}$} & \multicolumn{2}{c}{P-Value\,$^{\rm c}$} \\\cmidrule{2-9}
(vs $f_{\rm scatt}$) & \,\,\,\,\,\,Data\,$^{\rm d}$ & \,\,\,\,\,\,\,\,MC\,$^{\rm e}$ & \,\,\,\,\,Data\,$^{\rm d}$ & \,\,\,\,\,MC\,$^{\rm e}$ & \,\,\,\,\,Data\,$^{\rm d}$ & \,\,\,\,\,MC\,$^{\rm e}$ & Data\,$^{\rm d}$ & MC\,$^{\rm e}$\\
\hline
\vspace{1mm}
$N_{\rm H}$ & $-0.52\pm0.04$ & $-0.47\pm0.03$ & \,\,\,$12.1\pm1.0$ & \,\,$10.9\pm0.7$ & -0.63 & -0.50 & \,\,$1.9\times 10^{-28}$ & \,\,\,$1.5\times 10^{-25}$\\
\vspace{1mm}
$L_{\rm [OIII]}/L_{\rm 2-10}$ & \,\,\,\,\,$0.26\pm0.05$ & \,\,\,\,\,$0.33\pm0.02$ & \,\,\,\,\,\,$0.76\pm0.11$ & \,\,\,\,\,$0.79\pm 0.06$ & \,\,0.35 & \,\,0.36 & $2.5 \times 10^{-8}$ & \,\,\,$2.8\times10^{-12}$\\
\vspace{1mm}
$L_{\rm [OIII]}/L_{\rm 14-195}$ & \,\,\,\,\,$0.22\pm0.05$ & \,\,\,\,\,$0.27\pm0.02$ & \,\,\,\,\,\,$0.75\pm0.15$ & \,\,\,\,\,$0.76\pm0.07$ & \,\,0.27 & \,\,0.28 & $2.0\times10^{-5}$ & $9.0\times10^{-8}$\\
\vspace{1mm}
$L^{\rm corr}_{\rm [OIII]}/L_{\rm 2-10}$ & \,\,\,\,\,$0.27\pm0.05$ & \,\,\,\,\,$0.36\pm0.03$ & \,\,\,\,\,\,$0.65\pm0.10$ & \,\,\,\,\,$0.70\pm0.05$ & \,\,0.33 & \,\,0.34 & $4.9\times10^{-7}$ & \,\,\,$3.2\times10^{-10}$\\
\vspace{1mm}
$L^{\rm corr}_{\rm [OIII]}/L_{\rm 14-195}$ & \,\,\,\,\,$0.20\pm0.06$ & \,\,\,\,\,$0.27\pm0.03$ & \,\,\,\,\,\,$0.62\pm0.14$ & \,\,\,\,\,$0.66\pm0.07$ & \,\,0.23 & \,\,0.24 & $6.0\times10^{-4}$ & $1.4\times10^{-5}$\\
\vspace{1mm}
$L_{\rm [OII]}/L_{\rm 2-10}$ & \,\,\,\,\,$0.31\pm0.05$ & \,\,\,\,\,$0.40\pm0.03$ & \,\,\,\,\,\,$1.01\pm0.16$ & \,\,\,\,\,$1.12\pm0.08$ & \,\,0.39 & \,\,0.38 & $1.4\times10^{-8}$ & \,\,\,$1.8\times10^{-11}$\\
\vspace{1mm}
$L_{\rm [OII]}/L_{\rm 14-195}$ & \,\,\,\,\,$0.26\pm0.06$ & \,\,\,\,\,$0.33\pm0.03$ & \,\,\,\,\,\,$0.97\pm0.20$ & \,\,\,\,\,$1.08\pm0.10$ & \,\,0.29 & \,\,0.29 & $2.7\times10^{-5}$ & $4.4\times10^{-7}$\\
$\lambda_{\rm Edd}$ & $-0.14\pm0.05$ & $-0.16\pm0.02$ & \,$-0.22\pm0.09$ & $-0.40\pm0.05$ & -0.23 & -0.20 & $2.1\times10^{-3}$ & $1.1\times10^{-3}$\\
\hline
\end{tabular}
\label{tab:results1}
\vspace{-2mm}
\begin{flushleft}$^{\rm a}$ Of the linear regression line.\\
$^{\rm b}$ The Pearson's correlation coefficient.\\
$^{\rm c}$ The probability of the data set appearing if the null hypothesis is correct.\\
$^{\rm d}$ Excluding upper limits.\\
$^{\rm e}$ \,Monte Carlo simulations.
\end{flushleft}
\end{table*}


\begin{table*}
\centering
\caption{Correlation results from MC simulations for Figure \ref{fig:fscatt_vs_logNH_Bin} (Section \ref{sect:corr1}).}
\begin{tabular}{ccccccc}
\hline
\hline
\,\,\,\,\,Parameter & Number of Sources & Bin & \,\,\,\,\,\,\,\,\,\,Slope\,$^{\rm a}$ & \,\,\,\,\,Intercept\,$^{\rm a}$ & \,\,\,\,\,\,\,\,R-Value\,$^{\rm b}$ & \,\,\,\,\,\,P-Value\,$^{\rm c}$ \\
\hline
\vspace{1mm}
\multirow{2}{*}{$L_{\rm 14-195}$} & 191/382 & $< 4.74\times10^{43} {\rm erg/s}$ & $-0.40\pm0.04$ & $\,\,\,9.2\pm0.9$ & $-0.43\pm0.04$ & \,\,\,$8.5\times10^{-10}$\\
 & 191/382 & $> 4.74\times10^{43} {\rm erg/s}$ & $-0.54\pm0.04$ & $12.4\pm0.9$ & $-0.57\pm0.04$ & \,\,\,$1.2\times10^{-17}$\\
\hline
\vspace{1mm}
\multirow{2}{*}{$M_{\rm BH}$} & 136/273 & $< 10^{7.96}\,M_{\rm \odot}$ & $-0.36\pm0.05$ & \,\,\,$8.2\pm1.1$ & $-0.42\pm 0.05$ & $4.6\times10^{-7}$\\
 & 137/273 & $\geq 10^{7.96}\,M_{\rm \odot}$ & $-0.42\pm0.05$ & \,\,\,$9.7\pm1.2$ & $-0.43\pm 0.05$ & \,$1.2\times10^{-7}$\\
\hline
\vspace{1mm}
\multirow{2}{*}{$\lambda_{\rm Edd}$} & 136/273 & $< 10^{-1.74}$ & $-0.41\pm0.05$ & \,\,\,$9.5\pm1.2$ & $-0.42\pm 0.05$ & $2.8\times10^{-7}$\\
 & 137/273 & $> 10^{-1.74}$ & $-0.32\pm0.05$ & \,\,\,$7.1\pm1.1$ & $-0.38\pm 0.05$ & $5.7\times10^{-6}$\\
\hline
\end{tabular}
\label{tab:results2}
\vspace{-2mm}
\begin{flushleft}$\,\,\,\,\,\,\,\,\,\,\,\,\,\,\,\,^{\rm a}$ Of the linear regression line.\\
$\,\,\,\,\,\,\,\,\,\,\,\,\,\,\,\,^{\rm b}$ The Pearson's correlation coefficient.\\
$\,\,\,\,\,\,\,\,\,\,\,\,\,\,\,\,^{\rm c}$ The probability of the data set appearing if the null hypothesis is correct.\\
\end{flushleft}
\end{table*}


We also used the simulated values of scattering fraction and column density to better quantify the correlation found between them. For each run of the MC simulations, we fitted a linear regression line to the simulated data and obtained the regression parameters such as slope, intercept, correlation coefficient, and the probability of the null hypothesis. The final slope, intercept, and Pearson's correlation coefficient (reported in Table \ref{tab:results1}) are calculated from the median of these 10,000 values, and the corresponding regression line is shown as a dashed red line in Figure \ref{fig:fscatt_vs_logNH}. Compared to the linear regression line fitted initially to the data after excluding the upper limits, the regression line found through MC simulations is more accurate since these simulations are a better representation of the range of values that the scattering fraction and the column density could take for each source. Hence, the corresponding regression parameters better define the observed negative correlation between $f_{\rm scatt}$ and ${\rm log}\,N_{\rm H}$. To confirm the reliability of this correlation, we examined the probability distribution of the linear regression obtained from MC simulations. As shown in Table \ref{tab:results1}, for the correlation between scattering fraction and column density, we get a probability $= 1.5\times10^{-25}$ for the null hypothesis. Such a low \textit{p}-value further confirms the significance of the inverse correlation obtained. We also fitted various higher degree polynomial regressions to the data points shown in Figure \ref{fig:fscatt_vs_logNH} but did not find any significant improvement in the chi-square values as compared to the one for the linear regression fit.

\begin{figure}
    \centering
    \includegraphics[width=0.49\textwidth]{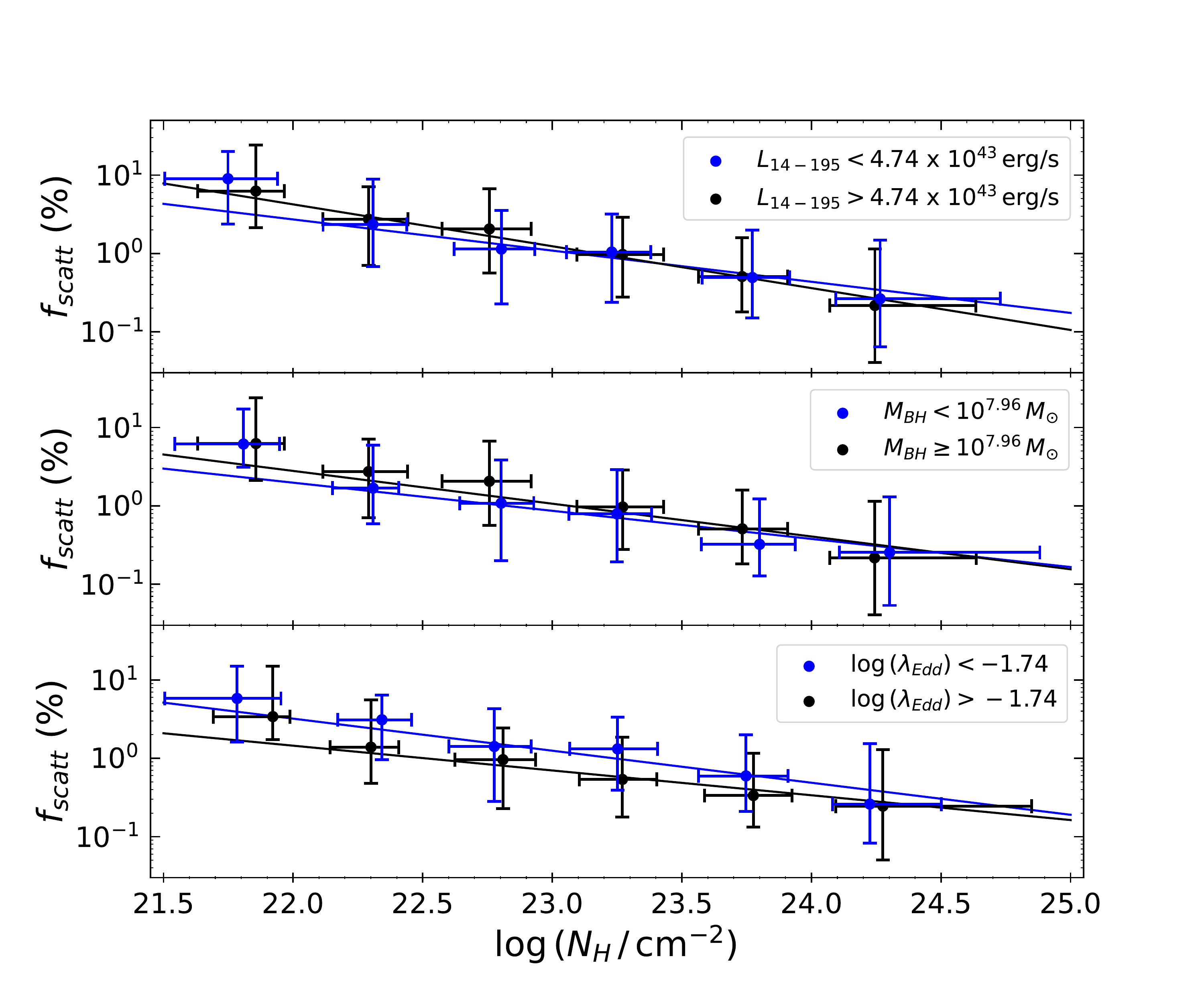}
    \caption{Scattering fraction vs column density for objects corresponding to two bins in: (a) Top panel: intrinsic 14--195 keV luminosity ($L_{\rm 14-195}$), (b) Middle panel: black hole mass ($M_{\rm BH}$), and (c) Bottom panel: Eddington ratio ($\lambda_{\rm Edd}$). The linear regression line as well as the median $f_{\rm scatt}$ and ${\rm log}\,N_{\rm H}$ values with 1$\sigma$ uncertainty for each ${\rm log}\,N_{\rm H}$ bin are calculated using MC simulations. The plots show a distinct negative correlation between scattering fraction and column density. The correlation parameters are given in Table \ref{tab:results2}.}
    \label{fig:fscatt_vs_logNH_Bin}
\end{figure}

We also check if the correlation we found between scattering fraction and column density holds for objects in different ranges of $L_{\rm 14-195}$, $M_{\rm BH}$, and $\lambda_{\rm Edd}$ in our sample. By specifying certain constraints on these physical properties of AGN, we aim to investigate how these external factors modify the correlation we found. To do so, we create two bins each of these three parameters, as follows: (a) $L_{\rm 14-195} < 4.74\times10^{43}$ erg/s and $> 4.74\times10^{43}$ erg/s, (b) $M_{\rm BH} < 10^{7.96} M_{\rm \odot}$ and $\geq 10^{7.96} M_{\rm \odot}$ and (c) ${\rm log}(\lambda_{\rm Edd}) < -1.74$ and $> -1.74$. These ranges were determined from the median values of these parameters. In Figure \ref{fig:fscatt_vs_logNH_Bin}, we show the linear regression line obtained using MC simulations for each of these intervals. We also show the median and uncertainty in the values of scattering fraction and column density for each bin calculated using MC simulations. The correlation parameters for these fits and the number of sources in each parameter window are reported in Table \ref{tab:results2}. Very low values of probability (of the null hypothesis) establish the significance of the negative correlation between $f_{\rm scatt}$ and ${\rm log}\,N_{\rm H}$ even when different conditions related to the physical properties of the accreting system are considered. Therefore, we can conclude that X-ray luminosity, black hole mass, and Eddington rate do not affect the negative correlation we found.


The X-ray spectra of 75/386 ($\sim 19\%$) sources in our sample were also fitted using torus models (see R17 for details). We have verified that including $f_{\rm scatt}$ and ${\rm log}\,N_{\rm H}$ values obtained from those fittings in our original data does not affect the anti-correlation found between these parameters. One should note that scattering fraction values $\geq 5\%-10\%$ can arise due to a partial covering absorber instead of Thomson scattering. To exclude their contributions, we apply an uppercut on the value of scattering fraction at 5$\%$ and 10$\%$. In Appendix \ref{sect:appendixa}, we show that applying such a criteria does not modify the correlation significantly (Figure \ref{fig:fscatt_vs_log_NH_FC}). The correlation we found may be biased by the typically low count rates of sources with high column densities. To make sure that these sources do not regulate the observed trend, we impose a lower limit on the number of counts per source at 200 counts. Figure \ref{fig:fscatt_vs_log_NH_CC} in Appendix \ref{sect:appendixa} shows that, even after removing the sources with low counts, we recover the same correlation between $f_{\rm scatt}$ and ${\rm log}\,N_{\rm H}$, using MC simulations. 


\subsection{Testing Parameter Degeneracies}\label{sect:deg}

An important step in our analysis is to check if the correlation we found is due to degeneracy between the two spectral parameters of interest, $f_{\rm scatt}$ and ${\rm log}\,N_{\rm H}$. Hence, in this section, we perform simulations to check for degeneracy between the scattering fraction and the column density. We simulated 100 dummy populations of obscured AGN, with the same size as our sample, which corresponds to more than 38,000 simulations, using a predefined model to describe the X-ray spectrum of an AGN. The model was decided based on its ability to provide a statistically significant fit to the X-ray spectra of the majority of sources in our sample. We used the same model to fit the simulated spectra and obtain simulated values of the parameters column density and scattering fraction. These two parameters were then plotted against each other to check if they show any inherent correlation and degeneracy.


\begin{figure*}
  \begin{subfigure}[t]{0.5\textwidth}
    \centering
    \includegraphics[width=\linewidth]{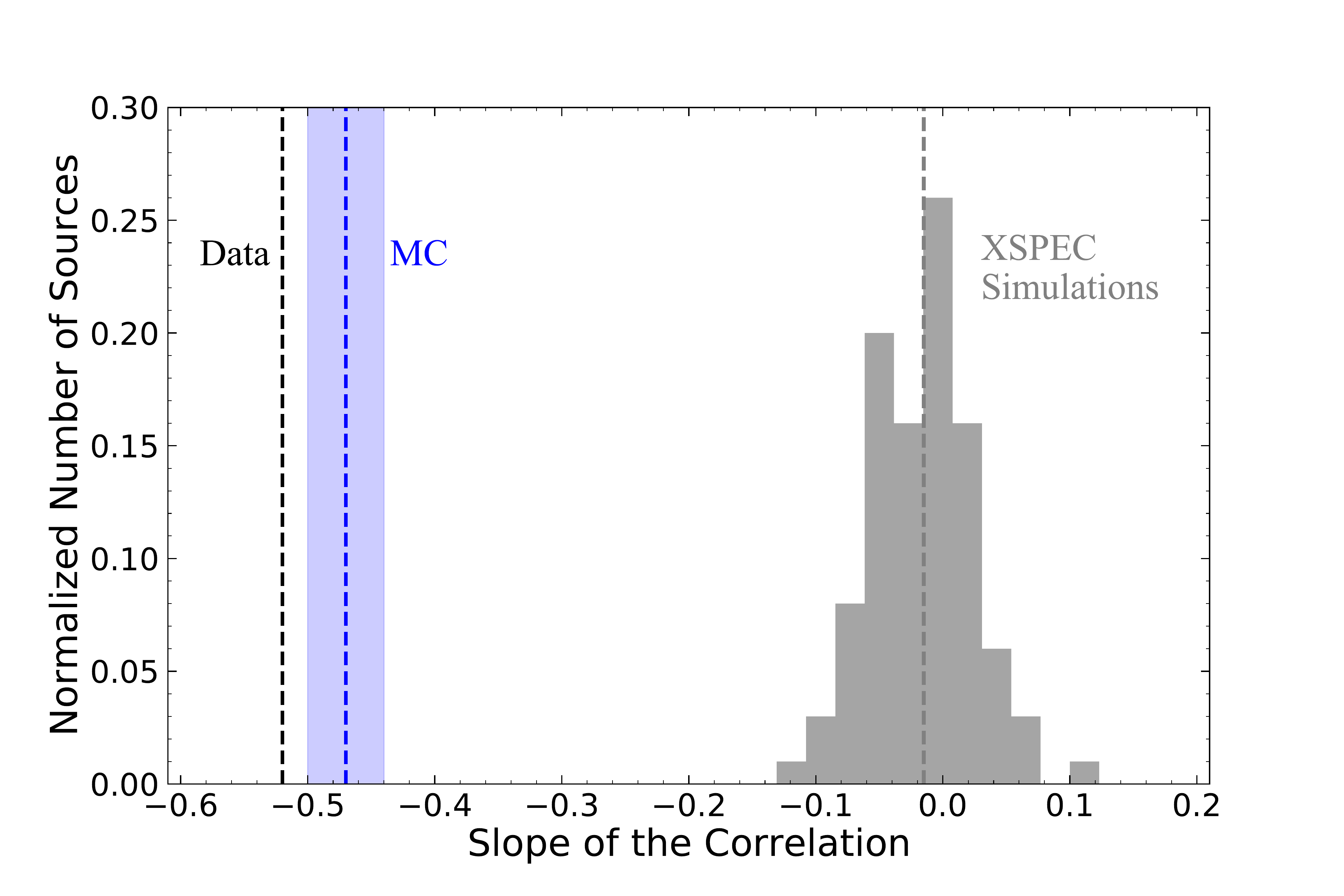}
    \caption{}
    \label{fig:slope}
  \end{subfigure}
  \hspace{-2mm}
  \begin{subfigure}[t]{0.5\textwidth}
    \centering
    \includegraphics[width=\linewidth]{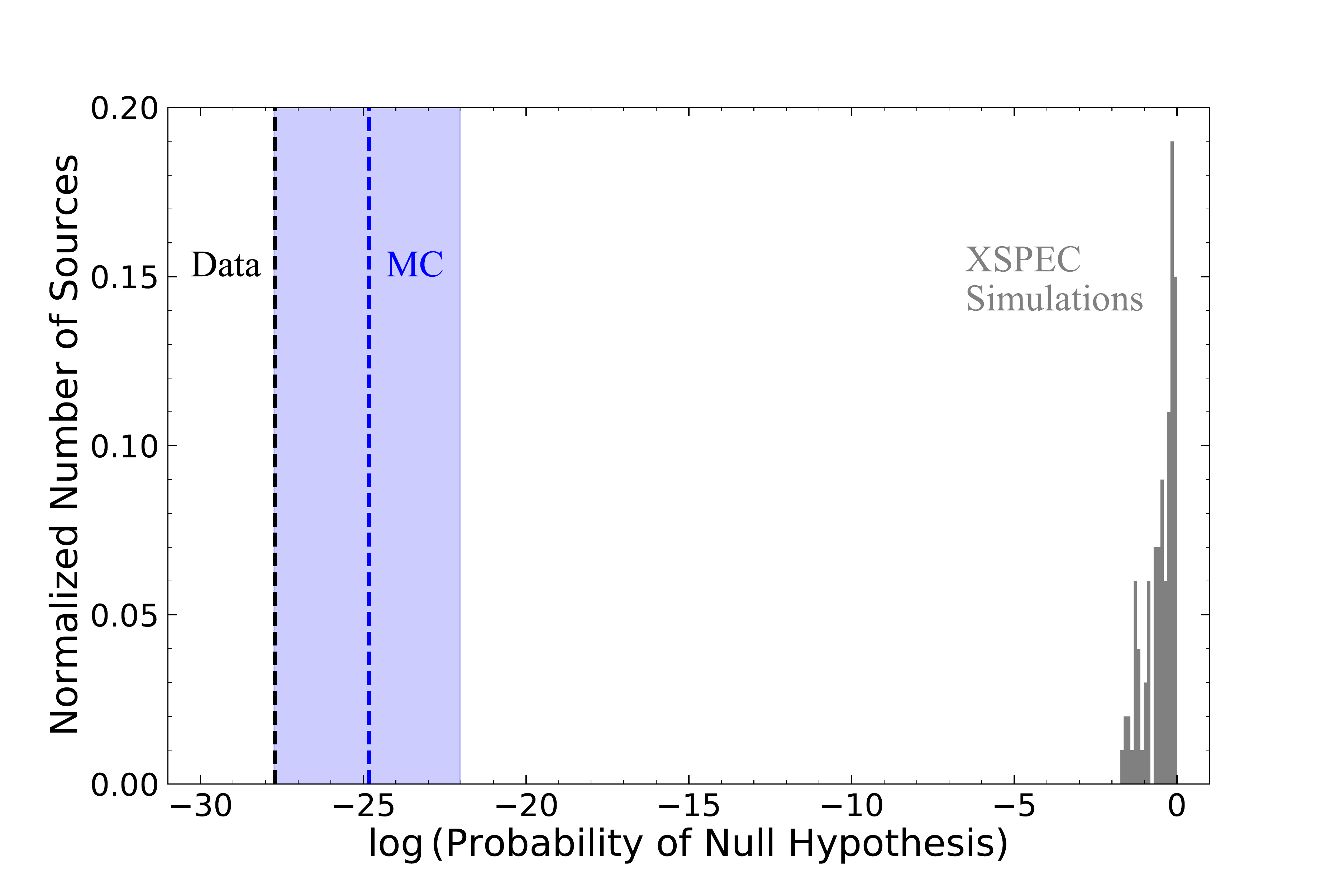}
    \caption{}
    \label{fig:prob}
  \end{subfigure}
  \caption{(a) Distribution in values of the slope of the correlation between scattering fraction and column density. (b) Distribution in the probability of the null hypothesis for the correlation between scattering fraction and column density.  The distribution obtained from the {\small XSPEC} simulations is shown in grey, with the median slope marked by a dashed grey line (in the left panel). The value obtained from the data (excluding upper limits) is shown as a dashed black line. The median value obtained from MC simulations (including upper limits) is shown by a dashed blue line with the shaded blue region showing the possible range of values. The difference in the values obtained from the {\small XSPEC} simulations and the data/MC simulations, for the slope of the correlation and the probability of the null hypothesis, shows that any degeneracy between the parameters ($f_{\rm scatt}$ and ${\rm log}\,N_{\rm H}$) does not produce an artificial correlation.}
 \label{fig:pd}
\end{figure*}


The X-ray spectra are simulated using the \texttt{FAKEIT} command in {\small XSPEC v12.10.1f} \citep{1996ASPC..101...17A}. As input, this command requires a background file, a response matrix file (RMF), and an auxiliary response file (ARF). For our analysis, we use the \textit{XMM-Newton}/EPIC-pn background, RMF, and ARF files. As a test, we repeat a set of simulations with \textit{Swift}/XRT background and response files and get similar results. We also redo the simulations with multiple sets of background and response files to accommodate any changes due to source positioning in the field of view and get similar results as to when we use a single set of background and response files. \texttt{FAKEIT} also needs an exposure time, which we randomly select from the distribution of the actual exposure times of the 386 sources in our sample. Finally, the \texttt{FAKEIT} command requires us to specify a model that will be used to generate fake spectra. For our simulations, we use the model B1 from R17 (described in detail in Section \ref{sect:xraydata}), which was used to fit the X-ray spectra of 272/386 ($\sim 71\%$) sources of our sample (see R17 for details). In {\small XSPEC} terminology, the model is defined by the following expression:\\

\texttt{ZPHABS} $\times$ \texttt{CABS} $\times$ \texttt{CUTOFFPL} $+$ \texttt{PEXRAV} $+$\\
\textit{constant} $\times$ \texttt{CUTOFFPL}\\

The various parameters in the model, their priors, and the conditions applied to them are explained in detail here:

\begin{itemize}
\item Scattering fraction ($f_{\rm scatt}$): This parameter is represented by the \textit{constant} term in the model. Its input value is selected randomly from a uniform distribution ranging from 0.1\% to 5\%.
\item Column density ($N_{\rm H}$): This parameter appears twice in the model (in components \texttt{ZPHABS} and \texttt{CABS}) and its value in these two components is tied. The default range acceptable for this parameter in the model needs to be modified to $ 0.1 < N_{\rm H}/10^{22}\,\rm cm^{-2} < 1000$ to describe the obscured AGN in our sample. The input value of this parameter is randomly selected from a uniform distribution of values between $10^{22}\,{\rm cm}^{-2}$ and $10^{24.5}\,{\rm cm}^{-2}$.
\item Photon index ($\Gamma$): This parameter appears three times in the model (in the two power-laws and the \texttt{PEXRAV} component), and its value in all three components is tied. The input value of $\Gamma$ is selected randomly from a Gaussian distribution centered around 1.8, with a standard deviation of 0.2 (\citealp{2017ApJS..233...17R}).
\item High energy cutoff ($E_{\rm C}$): This parameter also appears in the same three components as $\Gamma$, and its input value is selected randomly from a Gaussian distribution centered around 210 keV, with a standard deviation of 40 keV (\citealp{2018MNRAS.480.1819R}). The value of this parameter for the three components is tied and not allowed to vary.
\item Normalization ($K$): This parameter appears in the same three components as $\Gamma$ and $E_{\rm C}$, and its value in the three components is tied. It is calculated from flux values in the 2--10 keV band. Once the normalization values corresponding to all sources in our sample are calculated, the input value of this parameter in the model is selected randomly from the distribution of those values.
\item Reflection parameter ($R$): The default range acceptable for this parameter is set to $-1 < R < 0$, and its value is not allowed to vary. The input value is selected randomly from a Gaussian distribution centered around -0.4, with a standard deviation of 0.1 (\citealp{2017ApJS..233...17R}). The reflection parameter and luminosity could be anti-correlated (e.g. \citealp{2018ApJ...854...33Z}) with a preference for luminous AGN to have low to null reflection parameter (e.g. \citealp{1999ApJ...516..582V}; \citealp{2000MNRAS.316..234R}; \citealp{2005MNRAS.364..195P}). Furthermore, \cite{2011A&A...532A.102R} found different levels of reflection as a function of AGN type. However, we do not consider any such correlation between $R$ and $K$ or $R$ and $N_{\rm H}$ in these simulations. 
\item All the remaining parameters, such as redshift, iron abundance, and inclination angle, are set to their default values and not allowed to vary during the fitting process.
\end{itemize}

After feeding the model and the input values of all parameters to the software, we simulated 100 spectra each for all sources (total 38,600 spectra). All spectra were rebinned to have at least 20 counts per bin to use chi-squared statistics. We only generated the spectra in the 0.3--10 keV interval, matching the energy range covered by \textit{XMM-Newton}/EPIC-pn. All spectra were then fitted using the model defined above to obtain the best-fit values of all free parameters. For each set of simulated data, we then checked for the presence of a correlation between $f_{\rm scatt}$ and ${\rm log}\,N_{\rm H}$ with a linear fit, as it was done in Section \ref{sect:corr1} for our sample. To do so, we create a  distribution of the slopes of the linear regression lines for all 100 $f_{\rm scatt}$ vs ${\rm log}\,N_{\rm H}$ plots, as shown in Figure \ref{fig:slope}. The figure clearly shows a large difference between the slopes obtained from the data/MC simulations and the {\small XSPEC} simulations. Hence, we can safely infer from this distribution that the negative correlation we found between the scattering fraction and the column density for our sample of obscured AGN is not due to parameter degeneracy. We also show the probability distribution of the null hypothesis in Figure \ref{fig:prob}, which reinforces the validity of the inverse correlation we found. To confirm that these simulations do not miss sources with low $f_{\rm scatt}$ and low ${\rm log}\,N_{\rm H}$ values, that are otherwise not present in the original data (as visible in Figure \ref{fig:fscatt_vs_logNH}), we looked at the ratio of simulated $f_{\rm scatt}$ to input $f_{\rm scatt}$ as a function of input ${\rm log}\,N_{\rm H}$. The plot did not show any trend, hence, proving the ability of these simulations to recover such sources. An example of a simulated spectrum, along with the model used and the residuals after the fit, is shown in Figure \ref{fig:fake_spectrum} in Appendix \ref{sect:appendixb}. We also show an example of a simulated correlation between scattering fraction and column density values obtained from these simulations in Figure \ref{fig:fake_corr} in Appendix \ref{sect:appendixb}.

One should note that the scattered X-ray radiation could be absorbed by the host galaxy. To take this into account, we repeat a set of {\small XSPEC} simulations with a modified model including an absorbed scattered component. The column density of the absorbing material is assumed to be randomly distributed between $10^{19}\,{\rm cm}^{-2}$ and $10^{21}\,{\rm cm}^{-2}$. The simulated spectral parameters, column density and scattering fraction, hence obtained, do not appear to be correlated, thus confirming the lack of degeneracy between them. Since, throughout our simulations, we use the same model to create and fit the spectra, the validity of our approach to look for degeneracies may be debated. Therefore, we repeat a set of {\small XSPEC} simulations using two models, one to create the spectra and another to fit them. We employ the \texttt{RXTorus}\footnote{\textcolor{blue}{\url{https://www.astro.unige.ch/reflex/xspec-models}}} model developed by \cite{2017A&A...607A..31P} to simulate the fake spectra and the model B1 from R17 to fit them. Since we want to simulate the X-ray spectra of obscured sources, we assume an input value of $80^{\circ}$ for the viewing angle (where $0^{\circ}$ is face-on), and the ratio of the inner-to-outer radius of the torus is assumed to be 0.5. The input value of the equatorial column density is randomly selected from a uniform distribution of values between $10^{22}\,{\rm cm}^{-2}$ to $10^{24.5} \,{\rm cm}^{-2}$ and is connected to the line of sight column density via the inclination angle and the torus covering fraction. All the other parameters and their priors are the same as model B1 (described earlier). The resultant simulated parameters (${\rm log}\,N_{\rm H}$ and  $f_{\rm scatt}$) are then checked for any correlation using a linear regression fit, and we obtain a slope = $0.1\pm0.04$. Hence, we can conclusively confirm the absence of parameter degeneracy and the validity of the anti-correlation between the scattering fraction and column density. 


\section{The Correlation Between Scattering Fraction and $L_{\rm [O\,III]}/L_{\rm X}$, $L_{\rm [O\,III]}^{\rm corr}/L_{\rm X}$ and $L_{\rm [O\,II]}/L_{\rm X}$}\label{sect:corr2}

Having confirmed the negative correlation between the scattering fraction and the column density for our sample of obscured AGN, we now investigate if the scattering fraction relates with the ratio of [O{\scriptsize\,III}] $\lambda 5007$ to X-ray luminosity ($L_{\rm [O\,III]}/L_{\rm X}$). In the case of buried AGN with extremely low scattering fractions ($<0.5\%$), several studies have shown that these sources exhibit relatively lower values of $L_{\rm [O\,III]}/L_{\rm X}$ (e.g., \citealp{2010ApJ...711..144N}; \citealp{2015ApJ...815....1U}). As a further test, we also check for possible correlations between $f_{\rm scatt}$ and the ratio of [O{\scriptsize\,II}] $\lambda 3727$ to X-ray luminosity ($L_{\rm [O\,II]}/L_{\rm X}$), another tracer of the NLR. For our analysis, we adopt the observed [O{\scriptsize\,II}] $\lambda 3727$ luminosity ($L_{\rm [O\,II]}$) and the observed as well as extinction-corrected [O{\scriptsize\,III}] $\lambda 5007$ luminosity ($L_{\rm [O\,III]}$ and $L_{\rm [O\,III]}^{\rm corr}$, respectively). For $L_{\rm X}$, we use the intrinsic X-ray luminosity in the 2--10 keV and the 14--195 keV energy bands. Although, $L_{\rm 2-10}$ and $L_{\rm 14-195}$ are related quantities, $L_{\rm 2-10}$ can be strongly affected by high column densities. Therefore, we include the luminosity in the harder energy band (14--195 keV) because it is less biased by high column densities in our sample.


\begin{figure*}
  \begin{subfigure}[t]{0.5\textwidth}
    \centering
    \includegraphics[width=\textwidth]{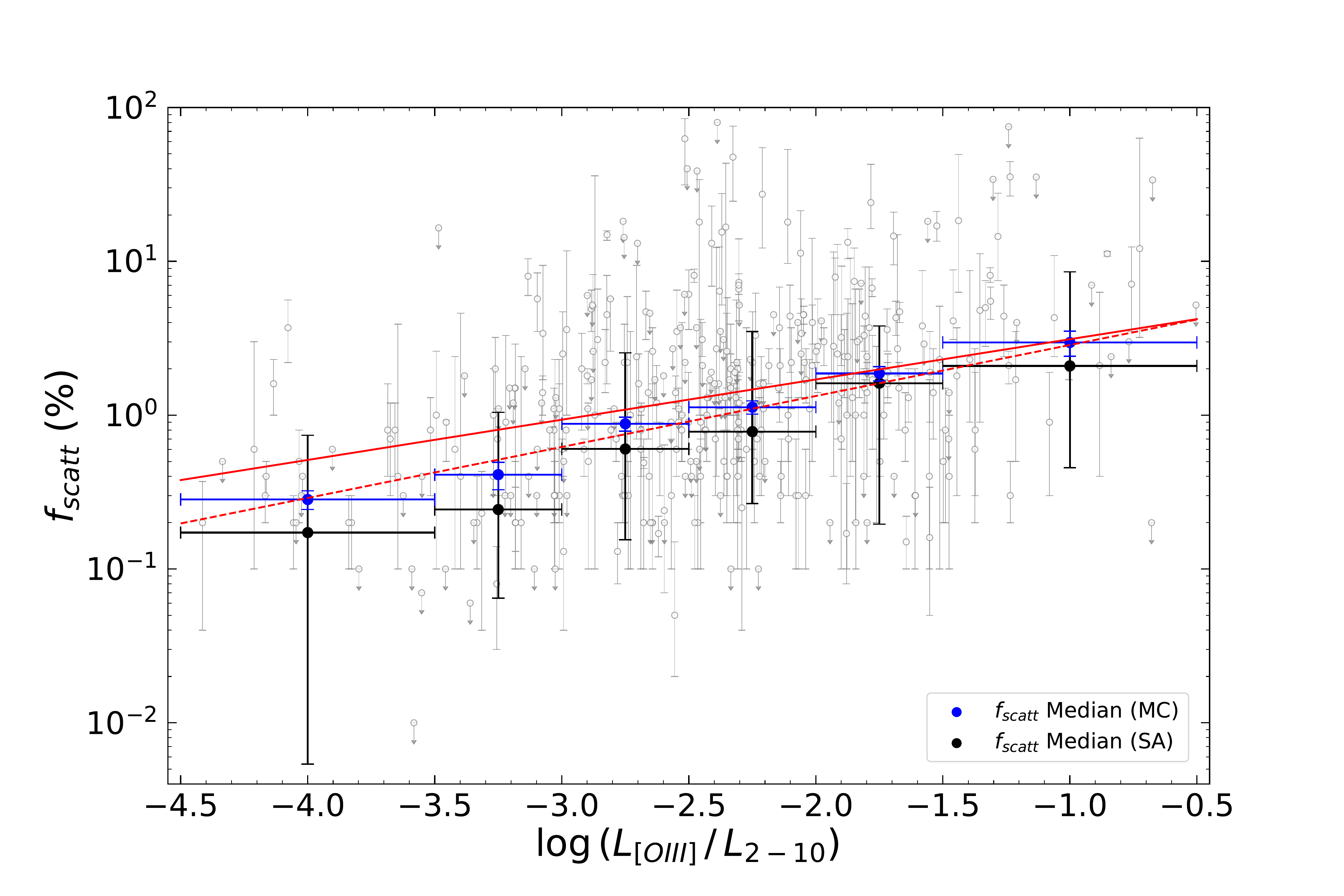} 
    \caption{}
    \label{fig:fscatt_vs_LOIII_L210}
  \end{subfigure}
  \hspace{-0.2cm}
  \begin{subfigure}[t]{0.5\textwidth}
    \centering
    \includegraphics[width=\textwidth]{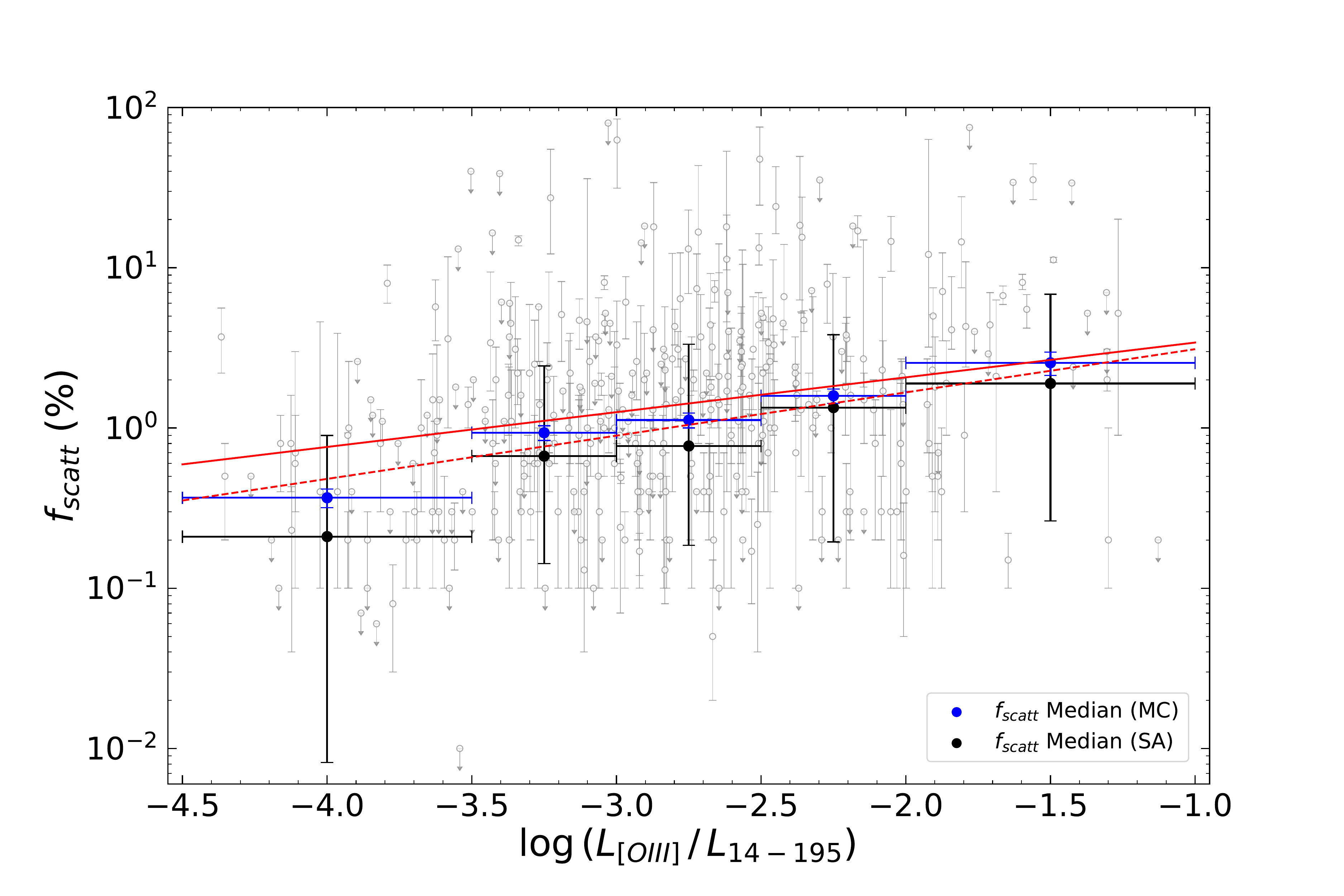}
    \caption{}
    \label{fig:fscatt_vs_LOIII_L14195}
  \end{subfigure}

  \begin{subfigure}[t]{0.5\textwidth}
    \centering
    \includegraphics[width=\textwidth]{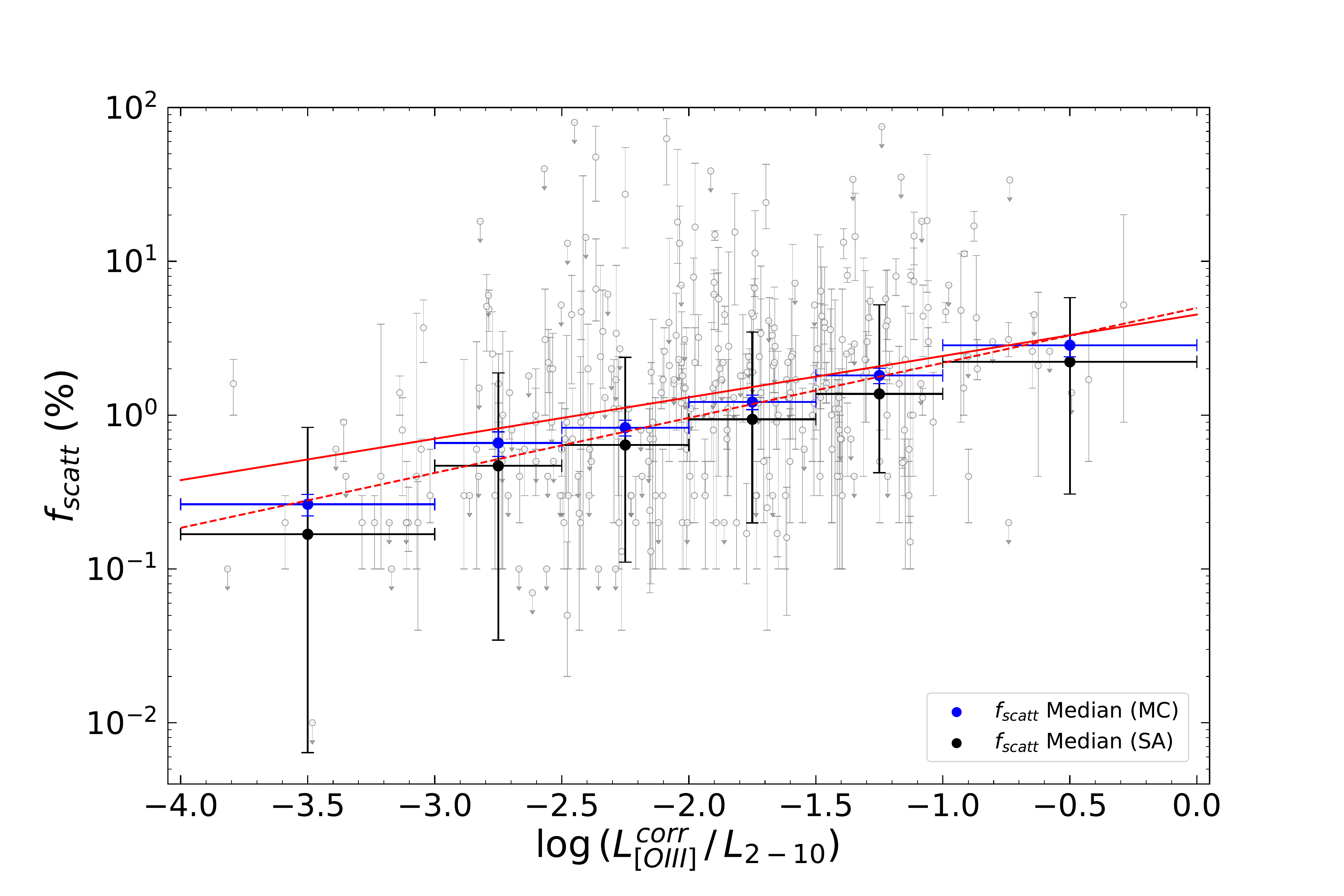}
    \caption{}
    \label{fig:fscatt_vs_LcOIII_L210}
  \end{subfigure}
  \hspace{-0.2cm}
  \begin{subfigure}[t]{0.5\textwidth}
    \centering
    \includegraphics[width=\textwidth]{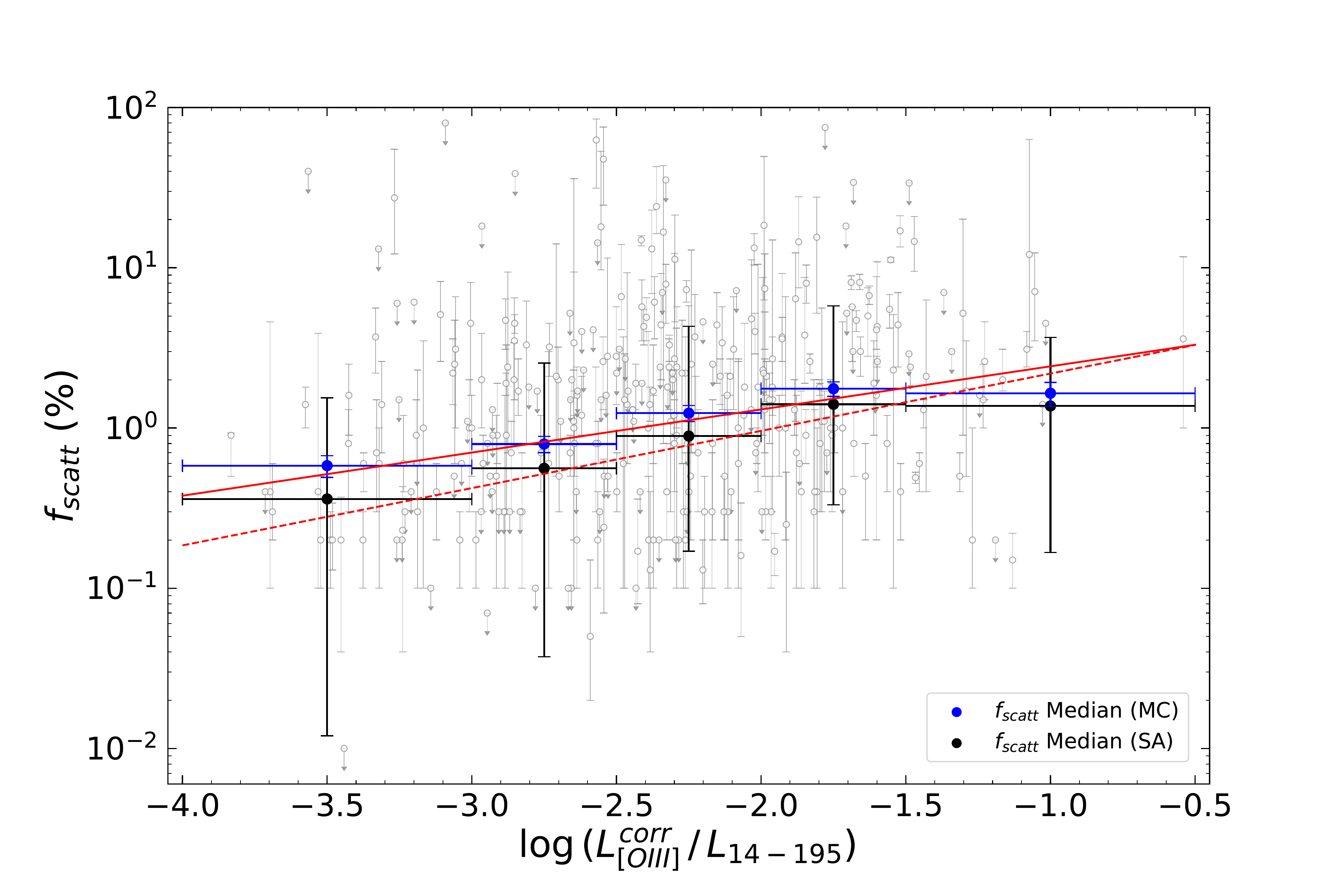}
    \caption{}
    \label{fig:fscatt_vs_LcOIII_L14195}
  \end{subfigure}

  \begin{subfigure}[t]{0.5\textwidth}
    \centering
    \includegraphics[width=\textwidth]{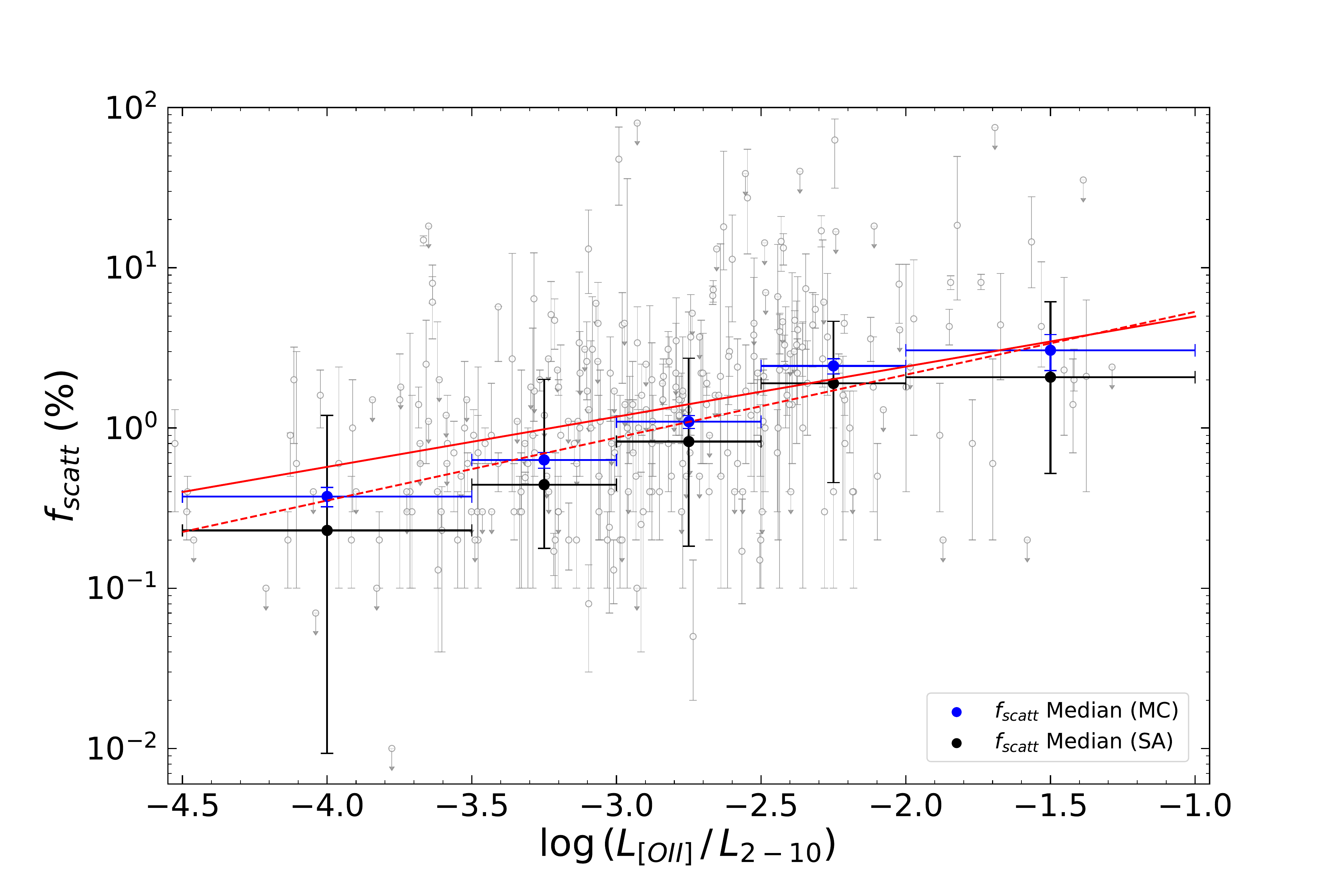}
    \caption{}
    \label{fig:fscatt_vs_LOII_L210}
  \end{subfigure}
  \hspace{-0.2cm}
  \begin{subfigure}[t]{0.49\textwidth}
    \centering
    \includegraphics[width=\textwidth]{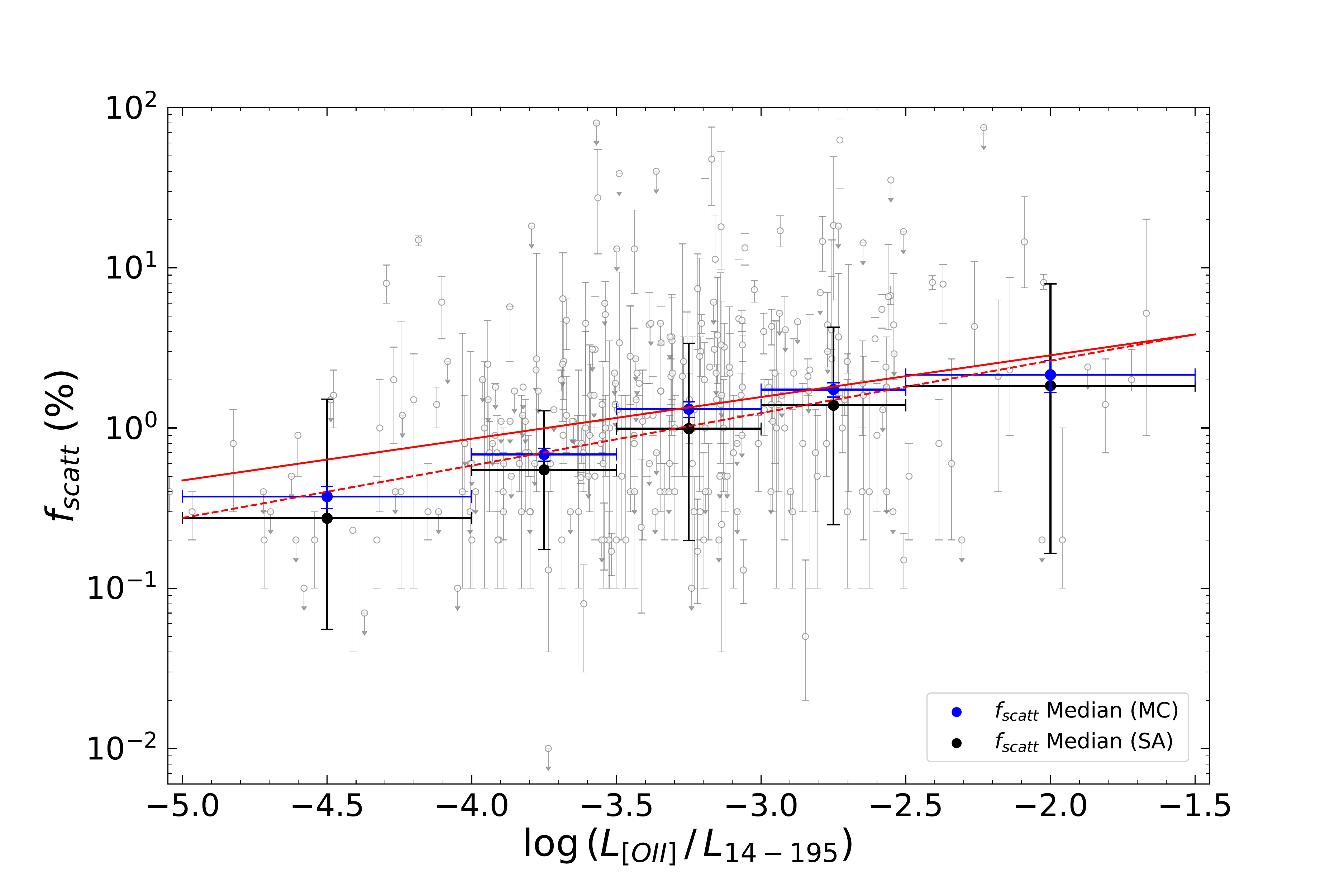}
    \caption{}
    \label{fig:fscatt_vs_LOII_L14195}
  \end{subfigure}
\caption{(a) Scattering fraction vs the ratio of observed [O{\scriptsize\,III}] $\lambda 5007$ to intrinsic 2--10 keV X-ray luminosity ($L_{\rm [OIII]}/L_{\rm 2-10}$). (b) Scattering fraction vs the ratio of observed [O{\scriptsize\,III}] $\lambda 5007$ to intrinsic 14--195 keV X-ray luminosity ($L_{\rm [OIII]}/L_{\rm 14-195}$). (c) Scattering fraction vs the ratio of extinction-corrected [O{\scriptsize\,III}] $\lambda 5007$ to intrinsic 2--10 keV X-ray luminosity ($L^{\rm corr}_{\rm [OIII]}/L_{\rm 2-10}$). (d) Scattering fraction vs the ratio of extinction-corrected [O{\scriptsize\,III}] $\lambda 5007$ to intrinsic 14--195 keV X-ray luminosity ($L^{\rm corr}_{\rm [OIII]}/L_{\rm 14-195}$). (e) Scattering fraction vs the ratio of observed [O{\scriptsize\,II}] $\lambda 3727$ to intrinsic 2-10 keV X-ray luminosity ($L_{\rm [OII]}/L_{\rm 12-10}$). (f) Scattering fraction vs the ratio of observed [O{\scriptsize\,II}] $\lambda 3727$ to intrinsic 14--195 keV X-ray luminosity ($L_{\rm [OIII]}/L_{\rm 14-195}$). Our data is shown as grey open circles in the background (upper limits as downward arrows and best-fit values with error bars showing errors at 90\% confidence level). The black and blue circles with error bars correspond to the median and 1$\sigma$ uncertainty in $f_{\rm scatt}$ calculated using SA and MC simulations, respectively. The solid red linear regression line is obtained from the data by excluding upper limits, while the dashed red regression line is obtained from MC simulations. The plots show a positive correlation between $f_{\rm scatt}$ and $L_{\rm [OIII]}/L_{\rm X}$, $L^{\rm corr}_{\rm [OIII]}/L_{\rm X}$ and $L_{\rm [OII]}/L_{\rm X}$. Correlation parameters are presented in Table \ref{tab:results1}.}  
\label{fig:fscatt_vs_LOIII}
\end{figure*}


Figure \ref{fig:fscatt_vs_LOIII_L210} shows $f_{\rm scatt}$ as a function of $L_{\rm [O\,III]}/L_{2-10}$ for 359 sources (in grey) for which we have the $L_{\rm [O\,III]}$ and $L_{\rm 2-10}$ measurements. Similar to the procedure followed in section \ref{sect:corr1}, we initially fit a linear regression line (in red) to all data points, excluding the upper limits. This fitting suggests that there might be a positive correlation between $f_{\rm scatt}$ and $L_{\rm [O\,III]}/L_{2-10}$ that needs to be analysed in more detail. To do so, we divide the data into six bins of $L_{\rm [O\,III]}/L_{2-10}$, such that each bin contains at least 20 sources. The median and standard error of $f_{\rm scatt}$ in each bin (shown in black), after including the upper limits, is calculated using the SA method. For the x-axis, we just plot the midpoint of each bin with error bars = 0.25 or 0.5, depending on the size of the bin. Furthermore, to get better estimates of the median and standard deviation in $f_{\rm scatt}$ in each bin, we use MC simulations. As described in detail in Section \ref{sect:corr1}, these simulations include the already calculated errors in the parameters to generate their probability distribution functions. However, it is important to note that we do not have error estimates for $L_{\rm [O\,III]}/L_{2-10}$. Therefore, in this case, we only created these probability distributions for $f_{\rm scatt}$. Whereas, for $L_{\rm [O\,III]}/L_{2-10}$, we just represent the midpoints and the size of each bin, as done for the SA method. Following the procedure in Section \ref{sect:corr1}, we obtained 10,000 values of scattering fraction for each source. For each run, depending on the $L_{\rm [O\,III]}/L_{2-10}$ bins, we classify the $f_{\rm scatt}$ values and calculate their median in each bin. The resulting median of each bin is the average of all 10,000 medians in that bin. The error in $f_{\rm scatt}$ in each bin is the standard deviation in the median values with respect to the final median (shown in blue). Therefore, these errors calculated from MC simulations are significantly smaller (by a factor of $\sqrt{N}$, where $N$ is the sample size in each bin) compared to those calculated from SA. Apart from this, the results from the two methods are consistent with each other. We also show the linear regression line calculated from the MC simulations (as a dashed line) in Figure \ref{fig:fscatt_vs_LOIII_L210}. All the correlation parameters are reported in Table \ref{tab:results1}.

In a similar manner as described above, the correlation between scattering fraction and $L_{\rm [O\,III]}/L_{14-195}$ is also inspected and is shown in Figure \ref{fig:fscatt_vs_LOIII_L14195}. It is clear from both plots (Figure \ref{fig:fscatt_vs_LOIII_L210} and \ref{fig:fscatt_vs_LOIII_L14195}) that scattering fraction correlates with $L_{\rm [O\,III]}/L_{\rm X}$ (where, X = 2--10 and 14--195). The significance of these two correlations is determined from their low $p$-values ($\approx 10^{-12}$ for $L_{\rm [O\,III]}/L_{2-10}$ and $\approx 10^{-8}$ for $L_{\rm [O\,III]}/L_{14-195}$). We also explore if this correlation persists when we correct the [O{\scriptsize\,III}] $\lambda 5007$ luminosities for reddening from the Balmer decrement (e.g., \citealp{2015ApJ...815....1U}). The relation between the scattering fraction and $L_{\rm [O\,III]}^{\rm corr}/L_{\rm X}$ for 318 sources is shown in Figure \ref{fig:fscatt_vs_LcOIII_L210} and \ref{fig:fscatt_vs_LcOIII_L14195}. The figures show a positive correlation similar to the one found between the scattering fraction and the ratio of the observed [O{\scriptsize\,III}] $\lambda 5007$ to X-ray luminosity. The null hypothesis probabilities for the correlation between $f_{\rm scatt}$ and $L^{\rm corr}_{\rm [O\,III]}/L_{\rm X}$ are quite low ($\approx 10^{-10}$ for X = 2--10 and $\approx 10^{-5}$ for X = 14--195; see Table \ref{tab:results1}). Thereby, highlighting the significance of the positive correlation between these two parameters. Finally, we examine the dependence of scattering fraction on the ratio of the observed [O{\scriptsize\,II}] $\lambda 3727$ to X-ray luminosity for 295 sources in our sample for which we could simultaneously obtain values of the observed [O{\scriptsize\,II}] $\lambda 3727$ luminosity and the intrinsic X-ray luminosity in the 2--10 and 14--195 keV bands. (Figure \ref{fig:fscatt_vs_LOII_L210} and \ref{fig:fscatt_vs_LOII_L14195}). As reported in Table \ref{tab:results1}, the correlations observed in this case have slightly higher slopes compared to those with $L_{\rm [O\,III]}/L_{\rm X}$. However, for both sets of correlations we get similar $p$-values ($<10^{-3}$), which confirms their significance. 

For the three sets of correlations evaluated in this section, it is worth noting that not always the narrow emission lines, such as [O{\scriptsize\,II}] $\lambda 3727$, [O{\scriptsize\,III}] $\lambda 5007$, H$\alpha$ and H$\beta$ can be ascribed solely to the AGN. In some AGN hosts, there might be a non-negligible contribution from star formation that cannot be ignored (e.g., \citealp{2018MNRAS.480.5203M}). To solve this issue, we checked the emission-line classification of all sources in our sample from \cite{2017ApJ...850...74K}. We used the [O{\scriptsize\,III}] $\lambda 5007$/H$\beta$ versus [N{\scriptsize\,II}] $\lambda 6583$/H$\alpha$ and [S{\scriptsize\,II}] $\lambda 6717$/H$\alpha$ classifications of \cite{1987ApJS...63..295V}, revised by \cite{2006MNRAS.372..961K}, to determine which sources lie in the star-forming regions of these line diagnostics diagrams. After removing those sources ($\sim15\%$), we recovered all the positive correlations between $f_{\rm scatt}$ and $L_{\rm [O\,III]}/L_{\rm X}$, $L^{\rm corr}_{\rm [O\,III]}/L_{\rm X}$, and $L_{\rm [O\,II]}/L_{\rm X}$.


\section{The Relation Between Scattering Fraction and $M_{\rm BH}$, $L_{\rm X}$ and $\lambda_{\rm Edd}$}\label{sect:corr3}

\begin{figure}
  \begin{subfigure}[t]{0.5\textwidth}
    \centering
    \includegraphics[width=\textwidth]{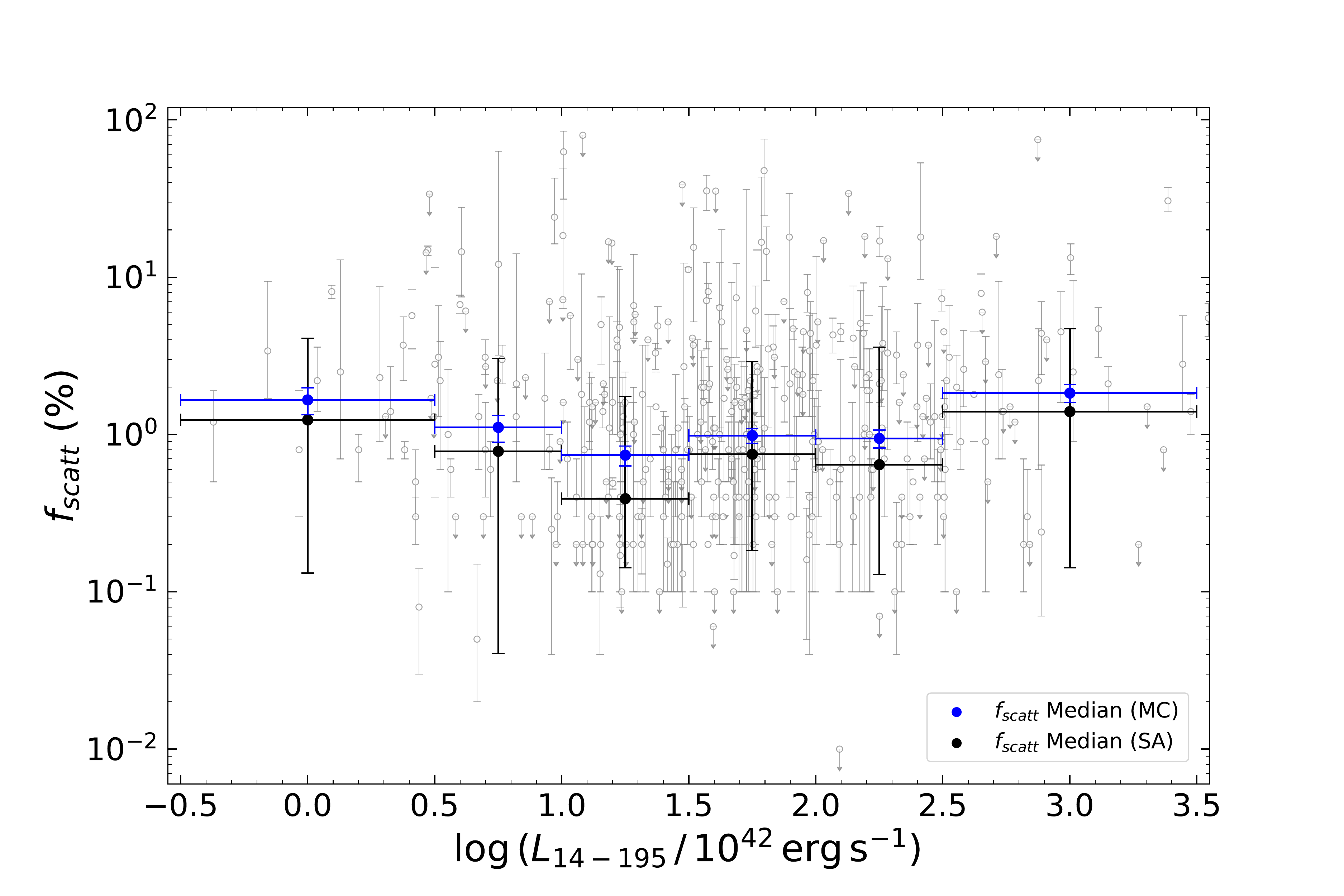}
    \caption{}
    \label{fig:fscatt_vs_L14_195}
  \end{subfigure}

  \begin{subfigure}[t]{0.5\textwidth}
    \centering
    \includegraphics[width=\textwidth]{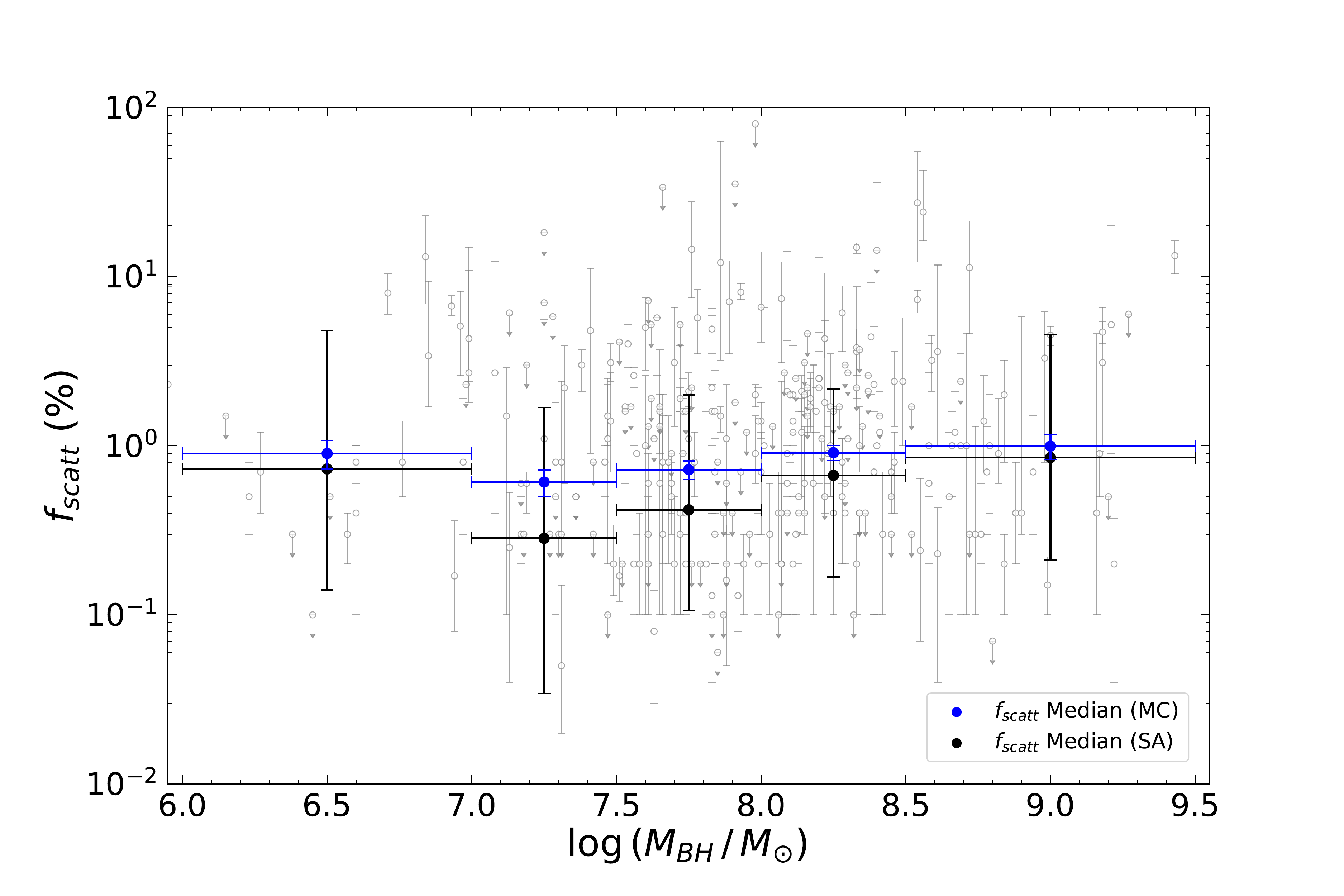}
    \caption{}
    \label{fig:fscatt_vs_M_BH}
  \end{subfigure}

  \begin{subfigure}[t]{0.5\textwidth}
    \centering
    \includegraphics[width=\textwidth]{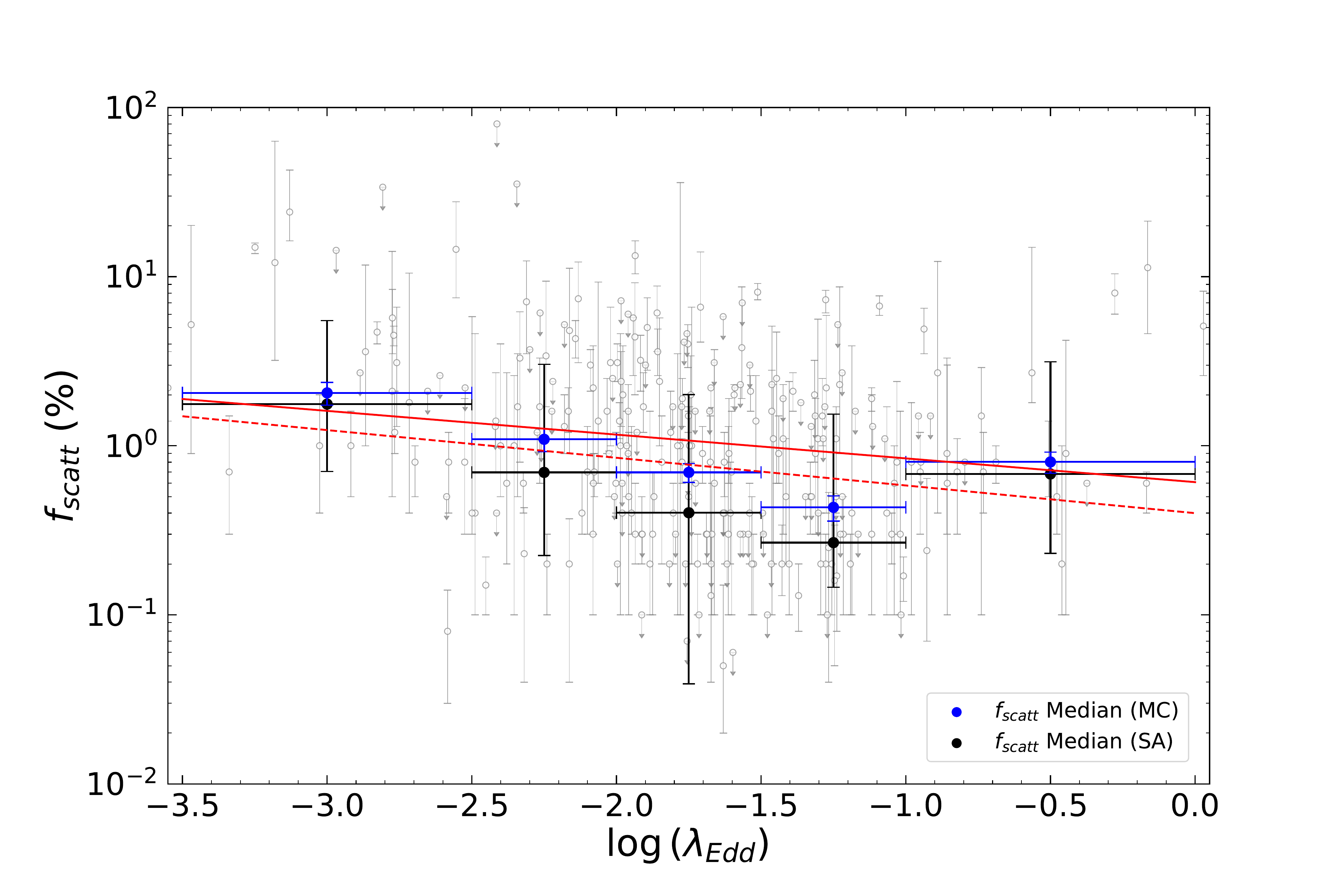}
    \caption{}
    \label{fig:fscatt_vs_Edd_Ratio}
  \end{subfigure}
\caption{(a) Scattering fraction vs intrinsic 14--195 keV luminosity. (b) Scattering fraction vs black hole mass. (c) Scattering fraction vs Eddington ratio. Our data is shown as grey open circles in the background (upper limits as downward arrows and best-fit values with error bars showing error at 90\% confidence level). The black and blue circles with error bars correspond to the median and 1$\sigma$ uncertainty in $f_{\rm scatt}$ calculated using SA and MC simulations, respectively. The solid red line (only in the bottom panel) depicts the linear regression fit to the data, excluding upper limits. The dashed red line (only in the bottom panel) is the regression fit obtained from MC simulations. The plots show no correlation between $f_{\rm scatt}$ and $L_{\rm 14-195}$ ($r$-value = 0.081) and $M_{\rm BH}$ ($r$-value = 0.071). A weak negative correlation is detected between $f_{\rm scatt}$ and $\lambda_{\rm Edd}$. Correlation parameters are reported in Table \ref{tab:results1}.} 
\label{fig:fscatt_vs_PP}
\end{figure}

In this section, we investigate how the scattering fraction evolves with different physical properties of AGN, such as X-ray luminosity, black hole mass, and Eddington ratio (Equation \ref{eq:1}). We have already seen that the correlation between scattering fraction and column density is not affected by any of these parameters. However, we want to check if the fraction of Thomson-scattered radiation is influenced by these properties. In Figure \ref{fig:fscatt_vs_L14_195}, we show the scattering fraction as a function of the intrinsic 14--195 keV luminosity ($L_{\rm 14-195}$), for 382 sources (in grey). Following the procedure described in Section \ref{sect:corr2}, we have divided the x-axis into six logarithmic bins of $L_{\rm 14-195}$ and plotted the median and error in $f_{\rm scatt}$, calculated using SA (in black) and MC simulations (in blue), for each bin. The plot shows no correlation ($p$-value = 0.11) between $f_{\rm scatt}$ and $L_{\rm 14-195}$. Figure \ref{fig:fscatt_vs_M_BH} shows scattering fraction as a function of black hole mass for 273 sources, for which we used black hole mass estimates from the BASS DR2 (Section \ref{sect:opticaldata}). The plot clearly shows a lack of correlation between $f_{\rm scatt}$ and $M_{\rm BH}$ ($p$-value = 0.24). 

Finally, we plot scattering fraction as a function of Eddington ratio for 273 sources in our sample (Figure \ref{fig:fscatt_vs_Edd_Ratio}). The Eddington ratio for these sources is calculated from Equation \ref{eq:1}, using the 2--10 keV bolometric correction from \cite{2009MNRAS.399.1553V} and the black hole masses. We find a weak negative correlation between $f_{\rm scatt}$ and $\lambda_{\rm Edd}$ ($p$-value $\approx 0.001$). Similar to the methods employed previously, we fit two linear regression lines to the data, one excluding the upper limits in $f_{\rm scatt}$ and the other including upper limits and errors in $f_{\rm scatt}$ using MC simulations. We have reported the regression parameters calculated from both methods in Table \ref{tab:results1}. \cite{2009MNRAS.392.1124V} showed that bolometric correction factors could be a function of the Eddington ratio. However, the correlation we found between $f_{\rm scatt}$ and $\lambda_{\rm Edd}$ does not change if we take into account such dependence.


\section{Discussion}\label{sect:discussion}

In the previous sections, we found various correlations between the scattering fraction and some physical properties of AGN, such as the line-of-sight column density, the ratio of the observed and extinction-corrected [O{\scriptsize\,III}] $\lambda 5007$ luminosity to X-ray luminosity, the ratio of the observed [O{\scriptsize\,II}] $\lambda 3727$ luminosity to X-ray luminosity, and Eddington ratio. In the following, we explore various interpretations of these correlations. In Section \ref{sect:torus}, we discuss two possible reasons behind the negative correlation between $f_{\rm scatt}$ and ${\rm log}\,N_{\rm H}$. In Section \ref{sect:nlr}, we discuss the dependence of $f_{\rm scatt}$ on $L_{\rm [O\,III]}/L_{\rm X}$ and other similar relations. We also explain how these correlations could shed light on the possible locus of Thomson scattering. And finally, in Section \ref{sect:edd}, we discuss about the weak negative correlation between $f_{\rm scatt}$ and $\lambda_{\rm Edd}$.


\subsection{The Effect of Inclination Angle and Torus Covering Factor on Scattering Fraction}\label{sect:torus}

For our sample of 386 hard-X-ray-selected, nearby, obscured AGN from the 70-month \textit{Swift}/BAT catalog, we find a negative correlation between the scattering fraction and the column density (Section \ref{sect:corr1} and Figure \ref{fig:fscatt_vs_logNH}). This trend is consistently observed even when we consider splitting the sample into different ranges of some of the fundamental parameters of accreting SMBHs, such as their X-ray luminosity, black hole mass, and Eddington ratio (Figure \ref{fig:fscatt_vs_logNH_Bin}). We exclude the possibility that this correlation is due to the degeneracy between different spectral parameters by simulating dummy populations of obscured AGN (Section \ref{sect:deg}). We fitted all 38,600 simulated spectra and explored possible correlations between $f_{\rm scatt}$ and $N_{\rm H}$. The slope and probability distributions (Figure \ref{fig:pd}) confirm the absence of any correlation between the simulated values, thereby implying that the correlation we found is intrinsic. Very high values of scattering fraction ($\geq 5\%-10\%$) can be either due to partially covering absorbers or due to significant contributions from radio jets (see \citealp{2017ApJS..233...17R} for a discussion). Therefore, we also check and confirm the existence of this correlation when an uppercut at $f_{\rm scatt}=5\%$ and $10\%$ is applied (Figure \ref{fig:fscatt_vs_log_NH_FC}). We also verify that this trend is not due to sources with low counts: by considering a lower limit of 200 counts per source, we recover the same correlation we find for the complete sample (Figure \ref{fig:fscatt_vs_log_NH_CC}).

We can, therefore, confirm the existence of a significant negative correlation ($p$-value $\approx 10^{-25}$) between the scattering fraction and the column density. We can also infer from our analysis that the dependence of the scattering fraction on the column density is not, in any way, affected by the X-ray luminosity of the accreting system, the mass of the central SMBH, or the Eddington ratio (Table \ref{tab:results2}). This implies that the main parameter driving the correlation is indeed the column density. To explain this inverse correlation between the scattering fraction and column density, we discuss here two possible explanations: (a) the inclination angle dependence of the Thomson cross-section and (b) the covering factor of the surrounding torus of AGN.


\begin{figure}
    \centering
    \includegraphics[width=0.48\textwidth]{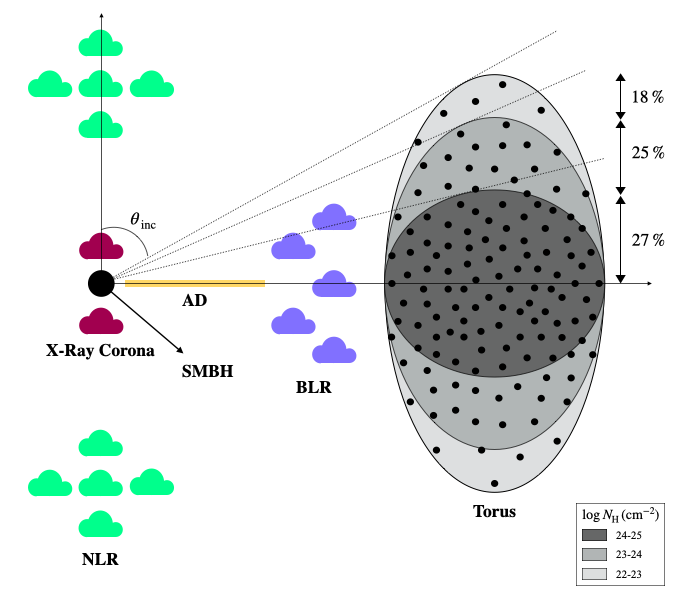}
    \caption{Schematic representation of an AGN showing a particular structure and geometry of the torus (also see Figure 4[b] of \citealp{2017NatAs...1..679R}). The inclination angle is measured from the normal to the accretion disk and relative to the observer. The intrinsic column density of the obscuring torus increases with inclination. The torus covering factor is determined from the fraction of sources within a specific column density range (details are given in \citealp{Ricci_2015} and \citealp{2017NatAs...1..679R}).}
    \label{fig:AGN}
\end{figure}


\subsubsection{\textbf{Inclination Angle}}\label{sect:angle}

To comprehend the reason behind the anti-correlation between scattering fraction and column density, we first consider the inclination angle dependence of the Thomson cross-section. The cross-section of Thomson scattering has a cosine square dependence on the inclination angle (e.g., \citealp{1986rpa..book.....R}). We also assume an average AGN model where the torus geometries and properties are similar for all sources, and the only parameter changing between the different sources is the inclination angle. A schematic of the assumed structure of the torus is shown in Figure \ref{fig:AGN}, where the column density increases with the inclination angle. We can also predict the covering factor ($f_{\rm cov}$) of different layers of the torus, based on the fraction of sources with line-of-sight column density in a specific range, following statistical arguments given by \cite{Ricci_2015} and \cite{2017NatAs...1..679R}. We start with a covering factor of 70\% for the torus, corresponding to a column density of $10^{22.5\pm0.5}\,{\rm cm}^{-2}$ and calculate the inclination angle ($\theta_{\rm inc}$) to be approximately $46^{\circ}$ (where, $\theta_{\rm inc}={\rm cos^{-1}}f_{\rm cov}$). As we consider higher values for the column density of the obscurer, at $10^{23.5\pm0.5}\,{\rm cm}^{-2}$ and $10^{24.5\pm0.5}\,{\rm cm}^{-2}$, the covering factor of the torus relative to that column density reduces to $52\%$ and $27\%$. Therefore, the viewing angles corresponding to those column densities increase to $59^{\circ}$ and $74^{\circ}$. Since the fraction of Thomson-scattered radiation (denoted by $f_{\rm scatt}$) will be proportional to the cross-section of Thomson scattering ($[\frac{ d\sigma}{d\Omega}]_{\rm unpol} = \frac{r_{\rm o}^2}{2}\times[1+{\rm cos}^2\theta_{\rm inc}]$, where $r_{\rm o} = e^2/mc^2$), we used ${\rm cos}^2\theta_{\rm inc}$ as a proxy for the scattering fraction and plotted it against the column density of the torus. We obtained an inverse correlation between these parameters with a slope $= -0.42\pm0.07$. This slope is consistent with what we found from our data using the MC simulations ($-0.47\pm0.03$). Hence, we can conclude that the observed trend could be explained by the inclination angle dependence of the Thomson cross-section. Some studies have shown that the equivalent width of the [O{\scriptsize\,III}] $\lambda 5007$ line could be an indicator of inclination (e.g., \citealp{2011MNRAS.411.2223R}; \citealp{2017MNRAS.464..385B}; \citealp{2018A&A...617A..81V}). As a further test, we checked for a possible relation between scattering fraction and the equivalent width of the [O{\scriptsize\,III}] $\lambda 5007$ line and did not find them to be correlated with each other.


\subsubsection{\textbf{Torus Covering Factor}}\label{sect:cf}

Another possible explanation for the anti-correlation between $f_{\rm scatt}$ and ${\rm log}\,N_{\rm H}$ could be that sources with high column densities tend to have, on average, a higher covering factor of the torus (e.g., \citealp{2011A&A...532A.102R}  \citealp{2012ApJ...747L..33E}; \citealp{2016ApJ...819..166M}; \citealp{2018ApJ...853..146T}; \citealp{2019A&A...626A..40P}). This would reduce the fraction of Thomson-scattered radiation and result in lower values of scattering fraction. This interpretation also puts into perspective the conclusions made for the population of buried AGN identified by \cite{2007ApJ...664L..79U}. They suggested that one of the plausible explanations for such low values of scattering fraction ($<0.5\%$) would be the presence of an obscuring torus with a very high covering factor. However, we also need to keep in mind the inherent degeneracy between the torus covering factor and the amount of material available for Thomson scattering. Due to this degeneracy, a low scattering fraction can also arise due to the deficiency of circumnuclear material needed for the scattering to occur (e.g.,  \citealp{2007ApJ...664L..79U}).


\subsection{X-ray Scattering in the Narrow-Line Region}\label{sect:nlr}

In Section \ref{sect:corr2}, we detect a positive correlation between the scattering fraction and the ratio of the observed [O{\scriptsize\,III}] $\lambda5007$ luminosity to the intrinsic X-ray luminosity (Figure \ref{fig:fscatt_vs_LOIII_L210} and \ref{fig:fscatt_vs_LOIII_L14195}). This correlation also persists when the [O{\scriptsize\,III}] $\lambda5007$ luminosity is corrected for extinction (Figure \ref{fig:fscatt_vs_LcOIII_L210} and \ref{fig:fscatt_vs_LcOIII_L14195}). Therefore, suggesting that the soft scattered X-ray radiation observed in obscured AGN may originate in the narrow-line region (NLR). It is well-established that emission lines present in the soft X-ray spectrum of obscured AGN are produced in a gas photoionized by the central active nucleus, rather than a collisionally ionized gas (e.g., \citealp{2002ApJ...575..732K}; \citealp{2004MNRAS.350....1S}; \citealp{2007MNRAS.374.1290G}; \citealp{2010MNRAS.405..553B}; \citealp{2010A&A...515A..47N}; \citealp{2017A&A...600A.135B}). We also know that the NLR is composed of material produced by photoionization, based on which, \cite{2006A&A...448..499B} conducted a high-resolution spectral analysis of a sample of Seyfert 2 galaxies and found a striking similarity in the extension and the overall morphology of the soft X-ray and the [O\,{\scriptsize III}] $\lambda5007$ emission. They showed, using photoionization models, that it is possible to have a gas photoionized by the central AGN, extending over hundreds of parsec and producing both the [O\,{\scriptsize III}] $\lambda5007$ and the soft X-ray emission. The correlation we found further supports the claim that the soft X-rays and the [O\,{\scriptsize III}] $\lambda5007$ emission are produced by the same gas, i.e., the NLR. This correlation is also in agreement with previous results in the literature that found low values of $L_{\rm [O\,III]}/L_{\rm X}$ for sources with very low scattering fraction (e.g., \citealp{2010ApJ...711..144N}; \citealp{2015ApJ...815....1U}).

Compared to $L_{\rm [O\,III]}/L_{\rm X}$ (slope = $0.33\pm0.02$), we recover a slightly steeper correlation (slope = $0.40\pm0.03$) when we consider $L_{\rm [O\,II]}/L_{\rm X}$  (Figure \ref{fig:fscatt_vs_LOII_L210} and \ref{fig:fscatt_vs_LOII_L14195}). An even steeper dependence (slope = 0.98) is quoted by \cite{2016ApJS..225...14K} for the ratio of [O\,{\scriptsize IV}] $\lambda24.89\,\micron$ luminosity to 10--50 keV X-ray luminosity. Considering the fact that the [O\,{\scriptsize IV}] $\lambda24.89\,\micron$ emission is less extincted by dust, compared to the [O\,{\scriptsize III}] $\lambda5007$ emission, it is expected that the scattering fraction correlates more with its ratio to the X-ray luminosity. Consequently, the various correlations of $f_{\rm scatt}$ with ratios of optical line luminosities to X-ray luminosity further strengthen our conclusion that Thomson scattering occurs in the NLR. We also investigated if the correlations we found are affected by the different aperture sizes of the sources in our sample (Appendix \ref{sect:appendixc}). To do so, we checked if the $L_{\rm [O\,III]}/L_{\rm X}$ correlates with the redshift or the physical width of the slits used for [O\,{\scriptsize III}] $\lambda5007$ measurements (reported by \citealp{2017ApJ...850...74K} and Koss et al., in prep.). As shown in Figure \ref{fig:OIII} in Appendix \ref{sect:appendixc}, we do not find any significant correlation between $L_{\rm [O\,III]}/L_{\rm X}$ and redshift or $L_{\rm [O\,III]}/L_{\rm X}$ and slit sizes. Therefore, we can exclude any aperture effects on the trends we observe. As the column density is much better constrained compared to the scattering fraction, we also checked if the ratio of the [O\,{\scriptsize III}] $\lambda5007$ to X-ray luminosity is correlated with the column density. However, we only obtain a very weak negative correlation between ${\rm log}\,N_{\rm H}$ and $L_{\rm [O\,III]}/L_{\rm X}$, with a slope = $-0.20\pm0.01$ and $p$-value = $1.1\times10^{-4}$ for the 2--10 keV band, and a slope = $-0.12\pm0.01$ and $p$-value = $2.0\times10^{-2}$ for the 14--195 keV range. This weak correlation disappears when we consider two bins in $f_{\rm scatt}$ around median $f_{\rm scatt} = 1.3$. Hence, we can conclude that the ratio of the [O\,{\scriptsize III}] $\lambda5007$ to X-ray luminosity is indeed correlated with the scattering fraction and not with the column density.

The scatter visible in all plots of Figure \ref{fig:fscatt_vs_LOIII} could be attributed to different variability timescales of the X-rays and the optical emission lines (e.g., \citealp{1993ARA&A..31..717M}; \citealp{2015MNRAS.451.2517S}; \citealp{2017MNRAS.464.1466O}; \citealp{2019ApJ...883L..13I}). This correlation between scattering fraction and [O\,{\scriptsize III}] $\lambda5007$ to X-ray luminosity ratio could be useful in removing the large scatter in the $L_{\rm X}$ versus $L_{\rm [O\,III]}$ relation (e.g., \citealp{2015ApJ...815....1U}). Figure \ref{fig:fscatt_vs_LOIII_L210} shows that the ratio of the optical to X-ray bolometric corrections could change by a factor of $\sim$1000 since a change in scattering fraction from 0.1\% to 5\% causes a variation of almost three orders of magnitude in $L_{\rm[O\,III]}/L_{\rm X}$. As we show in Section \ref{sect:corr2} and \ref{sect:corr3}, it is $L_{\rm [O\,III]}$ which changes with $f_{\rm scatt}$ and not $L_{\rm X}$, therefore, the ratio between the [O\,{\scriptsize III}] $\lambda5007$ and the X-ray bolometric corrections for different objects could vary depending on their scattering fractions. However, exploring these effects in detail is beyond the scope of this paper.  Finally, given the fact that low $f_{\rm scatt}$ AGN, which have higher column densities, have typically low ionized optical line luminosity with respect to the X-ray luminosity, we would expect optical surveys, which rely on narrow emission lines to identify obscured AGN, could miss a significant fraction of the population of heavily obscured AGN.


\subsection{The Role of Eddington Ratio}\label{sect:edd}

For our sample, we find a weak negative correlation between the scattering fraction and the Eddington ratio (Section \ref{sect:corr3} and Figure \ref{fig:fscatt_vs_Edd_Ratio}). A possible interpretation of this correlation can be acquired from the inverse correlation between $f_{\rm scatt}$ and ${\rm log}\,N_{\rm H}$. We discuss in Section \ref{sect:torus} that the low scattering fractions could be, at least in part, due to a higher covering factor of the torus. This would imply that sources with high accretion rates having lower scattering fractions tend to be surrounded by a thicker torus. A similar argument was proposed by \cite{2010ApJ...711..144N}, who also found the Eddington ratio to be anti-correlated with the scattering fraction. Another potential explanation of this inverse correlation is explored by \cite{2017Natur.549..488R}, who presented a negative correlation between the covering factor of the Compton-thin material and $\lambda_{\rm Edd}$. As a result, we would expect the scattering fraction to also decrease with increasing accretion rates since Thomson scattering occurs in the Compton-thin circumnuclear material. This explanation is supported by the idea that the amount of material surrounding an AGN is regulated by the radiation pressure (e.g., \citealp{2006MNRAS.373L..16F}; \citealp{2008MNRAS.385L..43F}; \citealp{2009MNRAS.394L..89F}). Hence, it is possible that more rapidly accreting AGN tend to have a lower fraction of Thomson-scattered radiation, either due to removal of Compton-thin circumnuclear material by radiation pressure or due to a geometrically thick torus.


\section{Summary and Conclusion}\label{sect:summary}

In this work, we used the 70-month \textit{Swift}/BAT catalog to study the properties of Thomson-scattered X-ray radiation in a sample of 386 hard-X-ray-selected, obscured AGN in the local Universe. We used X-ray spectral parameters reported by \cite{2017ApJS..233...17R}, black hole masses estimated by \cite{2017ApJ...850...74K} and Koss et al. (in prep.), and narrow-line fluxes calculated by Oh et al. (in prep.) to investigate possible correlations between the fraction of Thomson-scattered radiation and other physical properties of SMBHs, such as the line-of-sight column density, X-ray luminosity, black hole mass, Eddington ratio, [O{\scriptsize\,III}] $\lambda$5007 to X-ray luminosity ratio, and [O{\scriptsize\,II}] $\lambda$3727 to X-ray luminosity ratio. Here, we summarize our main findings:

\begin{itemize}
\item We found a significant negative correlation between the scattering fraction and the column density (see Section \ref{sect:corr1} and Figure \ref{fig:fscatt_vs_logNH}), with a slope $= -0.47\,\pm\,0.03$ (Table \ref{tab:results1}). We excluded the possibility that this correlation is due to sources with lower counts (Figure \ref{fig:fscatt_vs_log_NH_CC}) or due to partially covering absorbers ($f_{\rm scat}\geq5\%-10\%$; Figure \ref{fig:fscatt_vs_log_NH_FC}). We also verified that the correlation persists when considering different bins of X-ray luminosity, black hole mass, and accretion rate (Figure \ref{fig:fscatt_vs_logNH_Bin} and Table {\ref{tab:results2}}). 
\item To confirm that the correlation we found is intrinsic to our sample and not due to degeneracy between the parameters, we simulated and fitted more than 38,000 obscured AGN spectra (see Section \ref{sect:deg} and Figure \ref{fig:fake_spectrum}). We then checked for possible correlations between the simulated parameters. The probability and slope distribution plots (Figure \ref{fig:pd}) demonstrate that parameter degeneracy is not responsible for the observed trend (Figure \ref{fig:fake_corr}).
\item To understand the physical mechanism responsible for the anti-correlation between scattering fraction and column density, we discussed two possible explanations. First, we considered the inclination angle dependence of the Thomson cross-section and assumed a certain average geometry of the circumnuclear material for all AGN (see Figure \ref{fig:AGN} and Section {\ref{sect:discussion}}). Based on these assumptions, we were able to explain the trend we observe. A second explanation for this trend is that, in general, sources with high line-of-sight column densities tend to have higher covering factors of the surrounding torus. As a result, the amount of Thomson-scattered radiation could decrease, in turn reducing the scattering fraction for these sources.
\item We found a positive correlation between the scattering fraction and the ratio of [O{\scriptsize\,III}] $\lambda$5007 to X-ray luminosity (see Section \ref{sect:corr2} and Figure {\ref{fig:fscatt_vs_LOIII}}), with a slope $=0.33\pm0.02\,\,(0.27\pm0.02)$ for the intrinsic 2--10 (14--195) keV X-ray luminosity. A similar, but slightly steeper, correlation was observed for the ratio of [O{\scriptsize\,II}] $\lambda$3727 to X-ray luminosity. Both the trends suggest that Thomson-scattered radiation could originate in the same region responsible for the optical [O{\scriptsize\,III}] $\lambda$5007 and [O{\scriptsize\,II}] $\lambda$3727  emission, i.e., the NLR. This conclusion is also supported by several previous studies (e.g., \citealp{2006A&A...448..499B}; \citealp{2015ApJ...815....1U}; \citealp{2016ApJS..225...14K}).
\item We did not find any dependence of the fraction of Thomson-scattered radiation on the X-ray luminosity or black hole mass of the AGN. However, we detected a weak negative correlation between scattering fraction and Eddington ratio (see Section \ref{sect:corr3} and Figure \ref{fig:fscatt_vs_PP}) with a slope = $-0.16\pm0.02$. This relation could imply that rapidly growing SMBHs are either surrounded by a thicker torus \citep{2010ApJ...711..144N} or have radiatively driven away the Compton-thin material needed for scattering \citep{2017Natur.549..488R}.
\end{itemize}


\section*{Acknowledgements}

We would like to thank the anonymous referee for the prompt and constructive report that strengthened this paper. We thank Matthew Temple for his valuable comments. This work made use of data from the NASA/IPAC Infrared Science Archive and NASA/IPAC Extragalactic Database (NED), which are operated by the Jet Propulsion Laboratory, California Institute of Technology, under contract with the National Aeronautics and Space Administration. This research has made use of data and/or software provided by the High Energy Astrophysics Science Archive Research Center (HEASARC), which is a service of the Astrophysics Science Division at NASA/GSFC and the High Energy Astrophysics Division of the Smithsonian Astrophysical Observatory. K.K.G. was supported by a 2018 grant from the ESO-Government of Chile Joint Committee. We acknowledge support from Fondecyt Iniciacion grant 11190831 (CR), ANID grants CATA-Basal AFB-170002 (FEB), FONDECYT Regular 1190818 and 1200495 (FEB), and Millennium Science Initiative ICN12\_009 (FEB). MK acknowledges support from NASA through ADAP award 80NSSC19K0749. KO acknowledges support from the National Research Foundation of Korea (NRF-2020R1C1C1005462). EP acknowledges financial support under ASI/INAF contract 2017-14-H.0. 


\section*{Data Availability}

The data underlying this article are available in \cite{2017ApJS..233...17R}, \cite{2017ApJ...850...74K}, Koss et al. (in prep.) and Oh et al. (in prep.). Additional data will be shared on a reasonable request to the corresponding author.

\clearpage

\appendix

\section{The $f_{\rm scatt}-N_{\rm H}$ Correlation with Different Cuts}\label{sect:appendixa}

\begin{figure*}
  \begin{subfigure}[t]{0.5\textwidth}
    \centering
    \includegraphics[width=\textwidth]{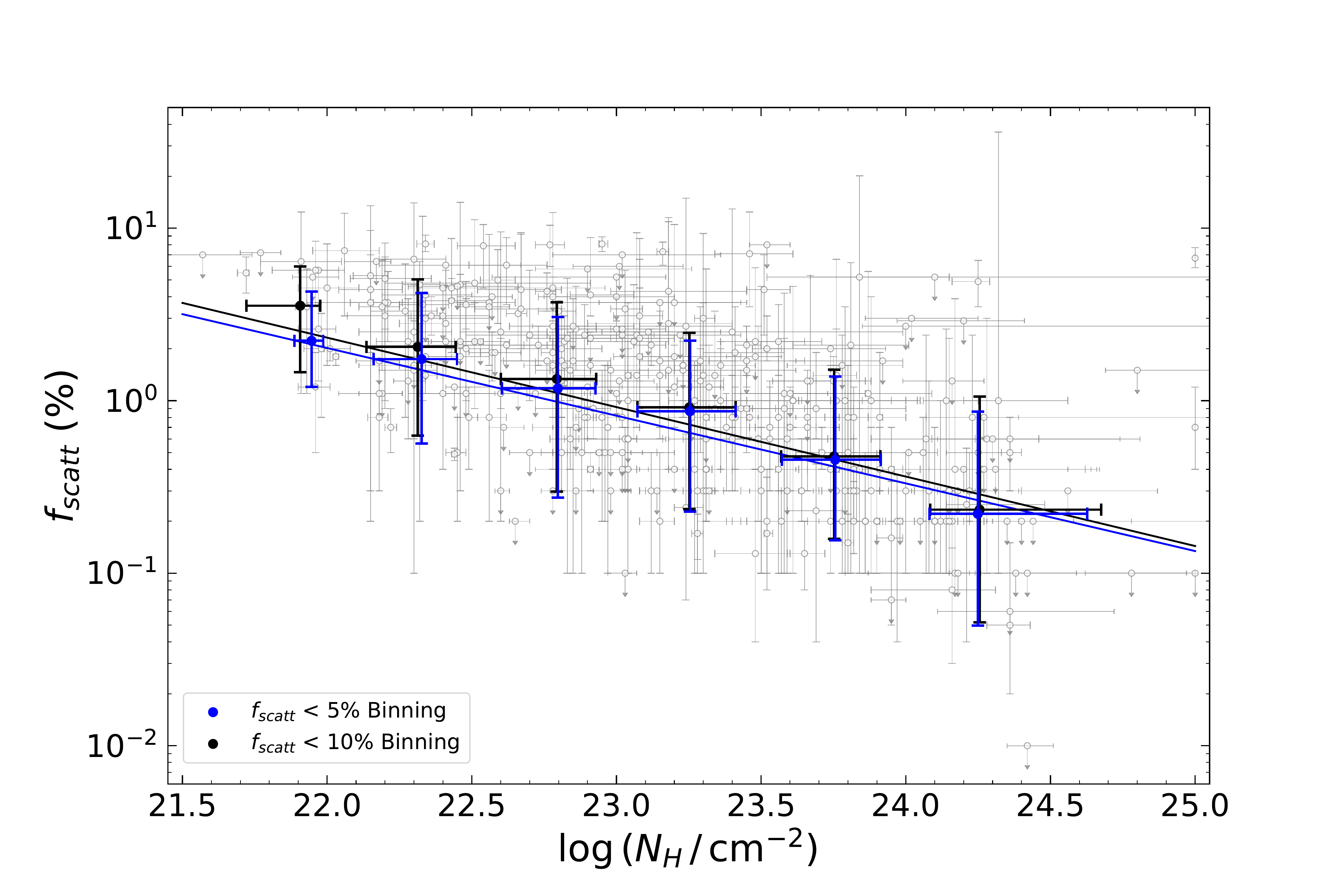} 
    \caption{}        
    \label{fig:fscatt_vs_log_NH_FC}
  \end{subfigure}
  \hspace{-0.2cm}
  \begin{subfigure}[t]{0.5\textwidth}
    \centering
    \includegraphics[width=\textwidth]{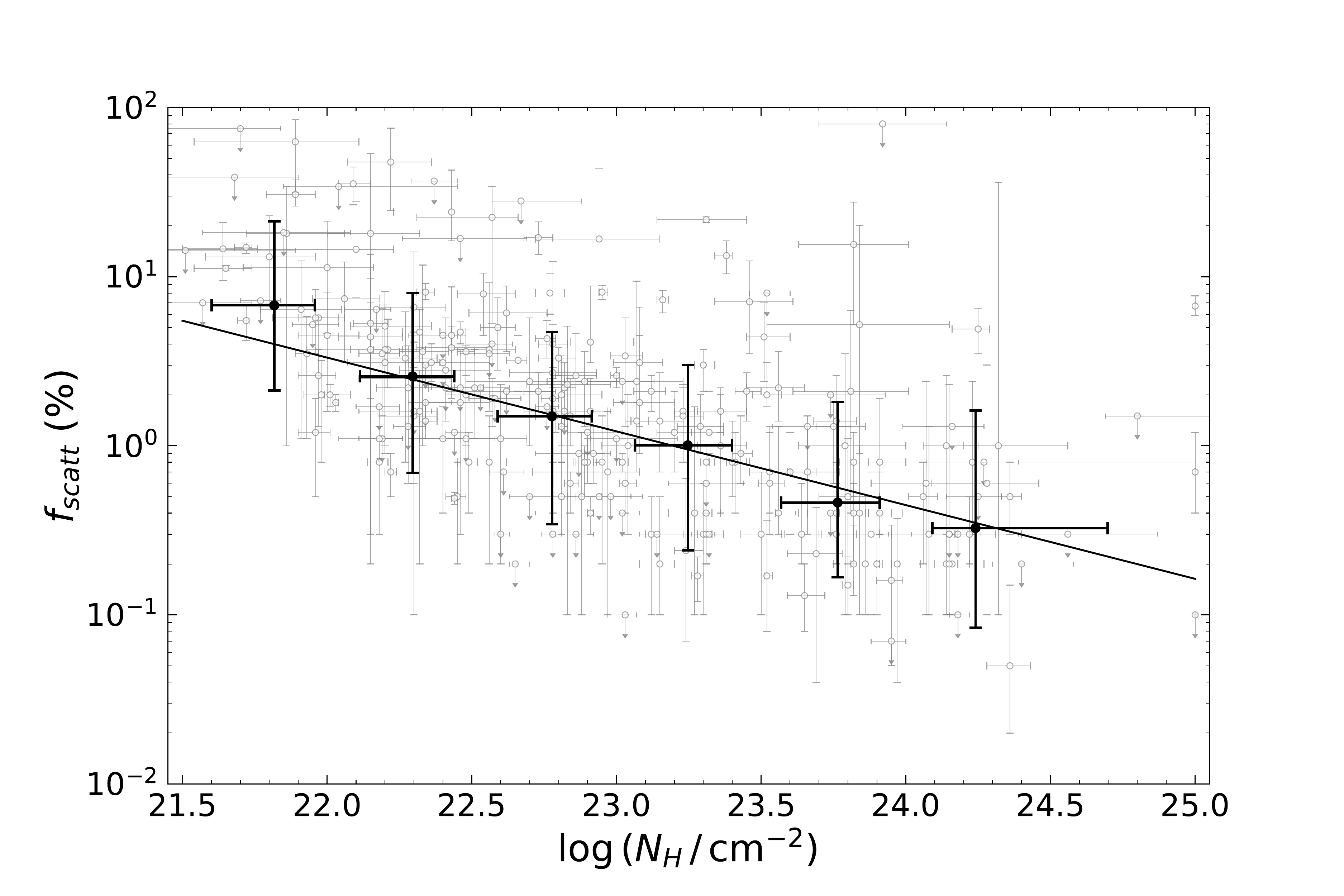}
    \caption{}
    \label{fig:fscatt_vs_log_NH_CC}
  \end{subfigure}
\caption{(a) Scattering fraction vs column density with an uppercut on $f_{\rm scatt}$ at 5\% (blue) and 10\% (black). The grey open circles in the background show the data points (with error bars showing errors at 90\% confidence level) for objects with $f_{\rm scatt}$ $<10\%$. The black and blue circles are the median and uncertainty (1$\sigma$) in $f_{\rm scatt}$ and ${\rm log}\,N_{\rm H}$ calculated using MC simulations for each ${\rm log}\,N_{\rm H}$ bin and both the linear regression lines are also obtained using the same method. (b) Scattering fraction vs column density for objects with $> 200$ counts. The data points for these objects are shown as grey open circles in the background (error bars show values in the 90\% confidence level). The black circles correspond to the median and uncertainty (1$\sigma$) in $f_{\rm scatt}$ and ${\rm log}\,N_{\rm H}$ calculated using MC simulations for each ${\rm log}\,N_{\rm H}$ bin, while the black line is the linear regression line obtained using the same method.}
\label{fig:fscatt_vs_logNH_APP1}
\end{figure*}

High values of scattering fraction ($\geq5\%-10\%$) in obscured AGN can be attributed to partially covering absorbers or to jet emission \citep{2017ApJS..233...17R}. To assure that these sources do not affect our analysis, we apply an uppercut on the values of scattering fraction at 5\% and 10\% and hence drop 67 and 38 sources, respectively. We plot the remaining sources as a function of column density in Figure \ref{fig:fscatt_vs_log_NH_FC}. Using MC simulations (see Section \ref{sect:corr1}), we calculate the correlation parameters and obtain a linear regression line with a slope $=-0.40\,\pm\,0.03$ for both cases, and \textit{p}-value $=6.5\times10^{-17}$ ($2.5\times10^{-19}$), for the 5\% (10\%) cut on $f_{\rm scatt}$. As evident from the plot, we find a trend similar to the one we obtain for the entire sample.

To verify that the anti-correlation we find between scattering fraction and column density is not due to sources with low counts, we apply a cut at 200 counts per source on our sample. It implies removing 136 sources from our original sample of 386 sources. We use MC simulations to quantify the dependence of $f_{\rm scatt}$ on ${\rm log}\,N_{\rm H}$ for the modified sample. Figure \ref{fig:fscatt_vs_log_NH_CC} demonstrates that these sources do not affect the original correlation, as we recover a similar correlation with a linear regression slope $=0.44\,\pm\,0.03$ and \textit{p}-value $=7.0\times10^{-16}$. 

\section{X-ray Spectral Simulations}\label{sect:appendixb}

\begin{figure*}
  \begin{subfigure}[t]{0.5\textwidth}
    \centering
    \includegraphics[width=\textwidth]{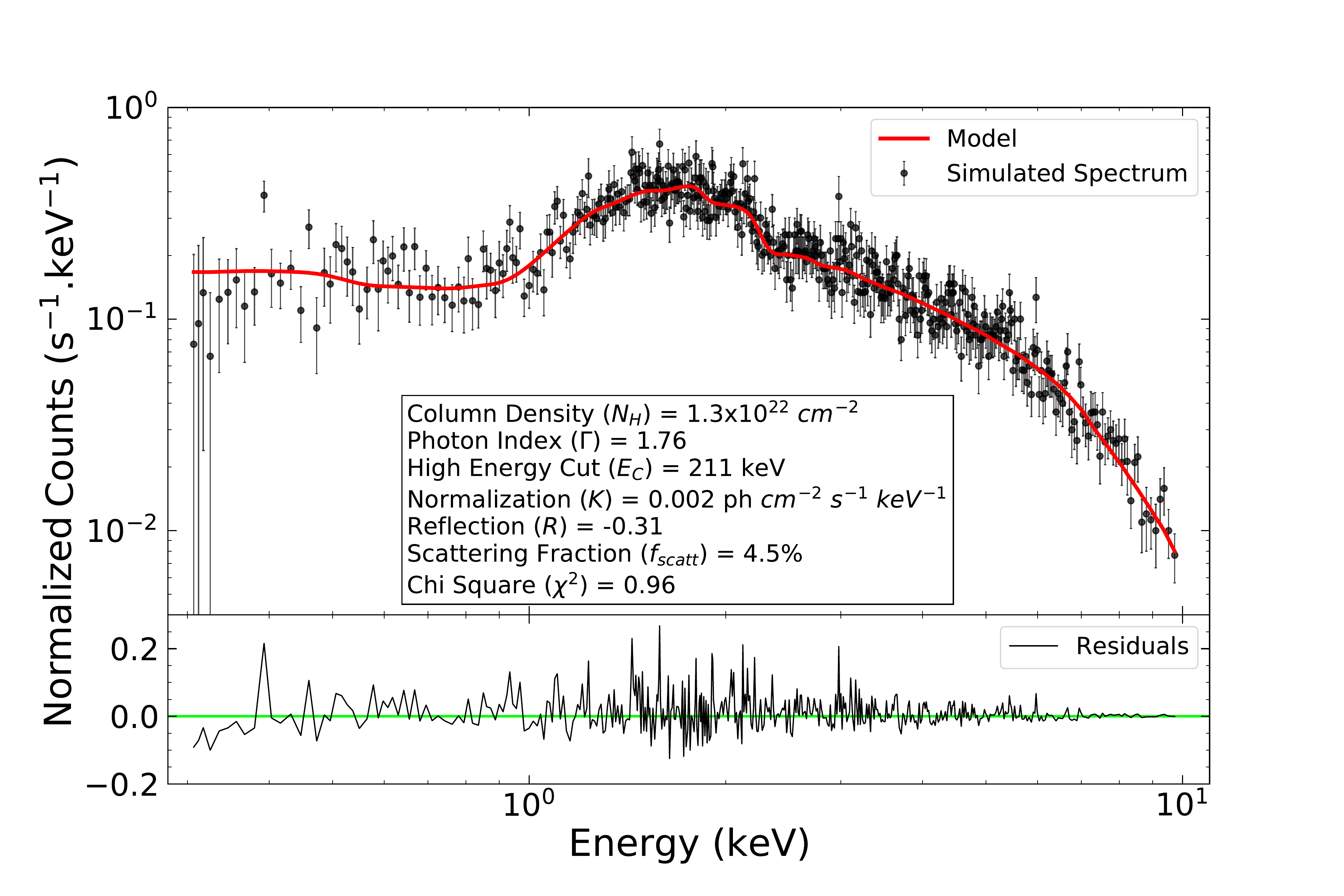} 
    \caption{}        
    \label{fig:fake_spectrum}
  \end{subfigure}
  \hspace{-0.2cm}
  \begin{subfigure}[t]{0.5\textwidth}
    \centering
    \includegraphics[width=\textwidth]{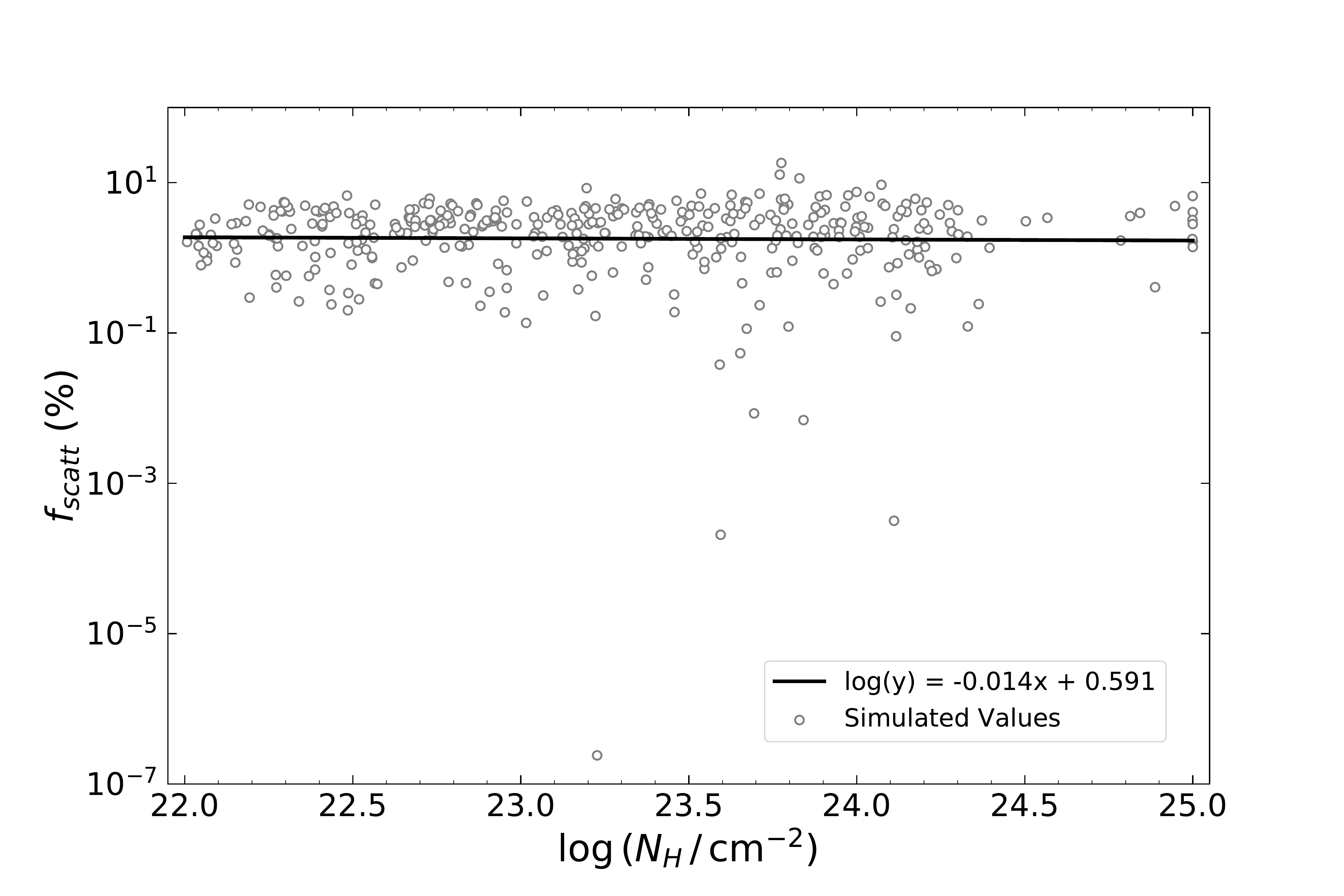}
    \caption{}
    \label{fig:fake_corr}
  \end{subfigure}
\caption{(a) An example of a simulated spectrum created using {\small XSPEC}. The model used to simulate and fit the spectrum is shown in red. The text box shows the best-fit values obtained for the various free parameters, while the bottom panel shows the fit residuals. (b) An example of a correlation between the simulated values of scattering fraction and column density obtained from {\small XSPEC} simulations. The black line shows the linear regression fit to the simulated dataset.}
\label{fig:fscatt_vs_logNH_APP2}
\end{figure*}

We ran multiple spectral simulations to exclude the possibility that the inverse correlation we find between the fraction of Thomson-scattered radiation and the line-of-sight column density is due to the degeneracy between these parameters (discussed in Section \ref{sect:deg}). In Figure \ref{fig:fake_spectrum}, we show an example of a simulated spectrum, along with the model used for simulating and fitting it and the residuals after the fitting. We also show in Figure \ref{fig:fake_corr} a correlation between the set of simulated parameters.


\section{Aperture Effects on [O{\scriptsize\,III}] $\lambda$5007 Measurements}\label{sect:appendixc}

\begin{figure*}
  \begin{subfigure}[t]{0.5\textwidth}
    \centering
    \includegraphics[width=\textwidth]{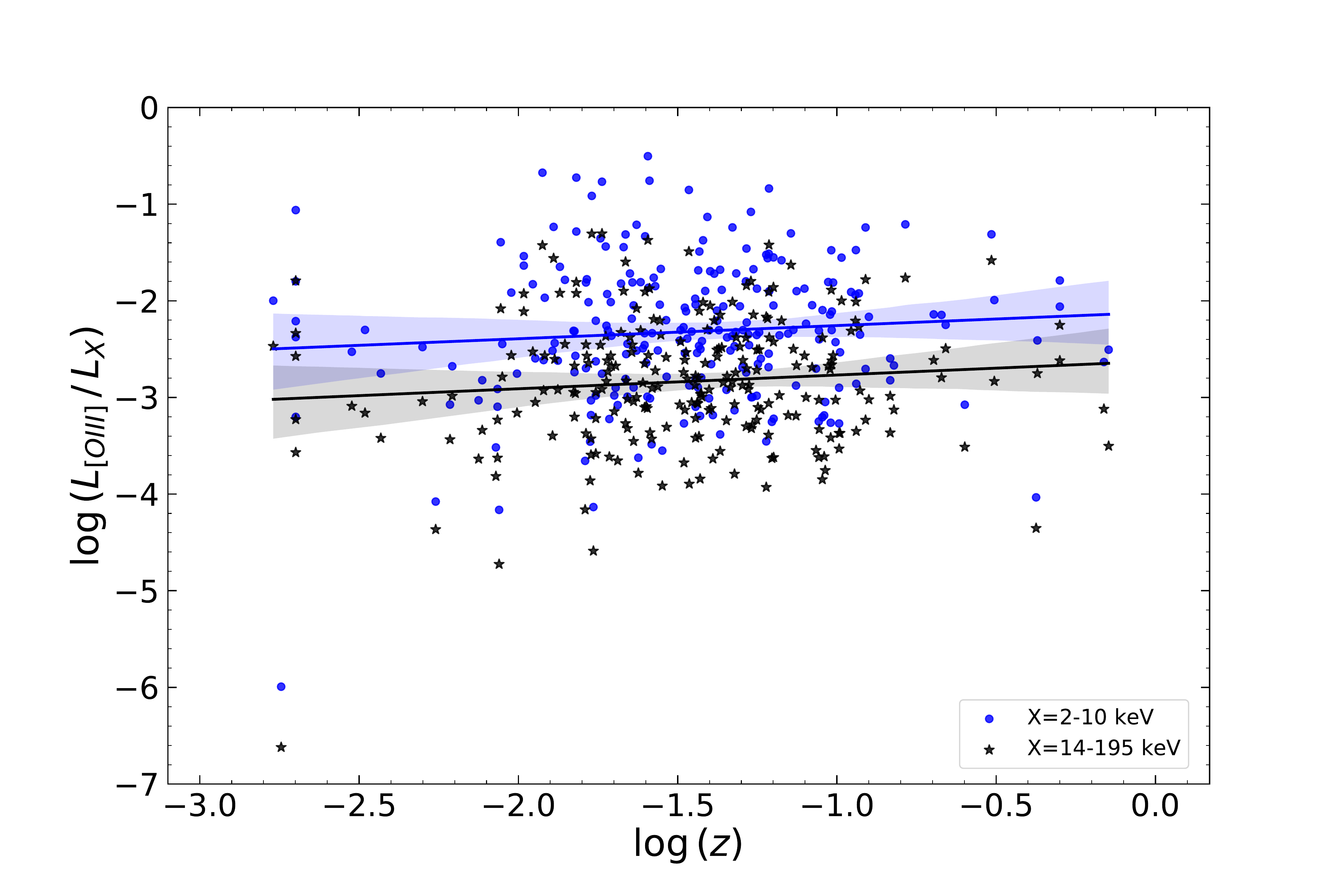} 
    \caption{}        
    \label{fig:OIIIz}
  \end{subfigure}
  \hspace{-0.2cm}
  \begin{subfigure}[t]{0.5\textwidth}
    \centering
    \includegraphics[width=\textwidth]{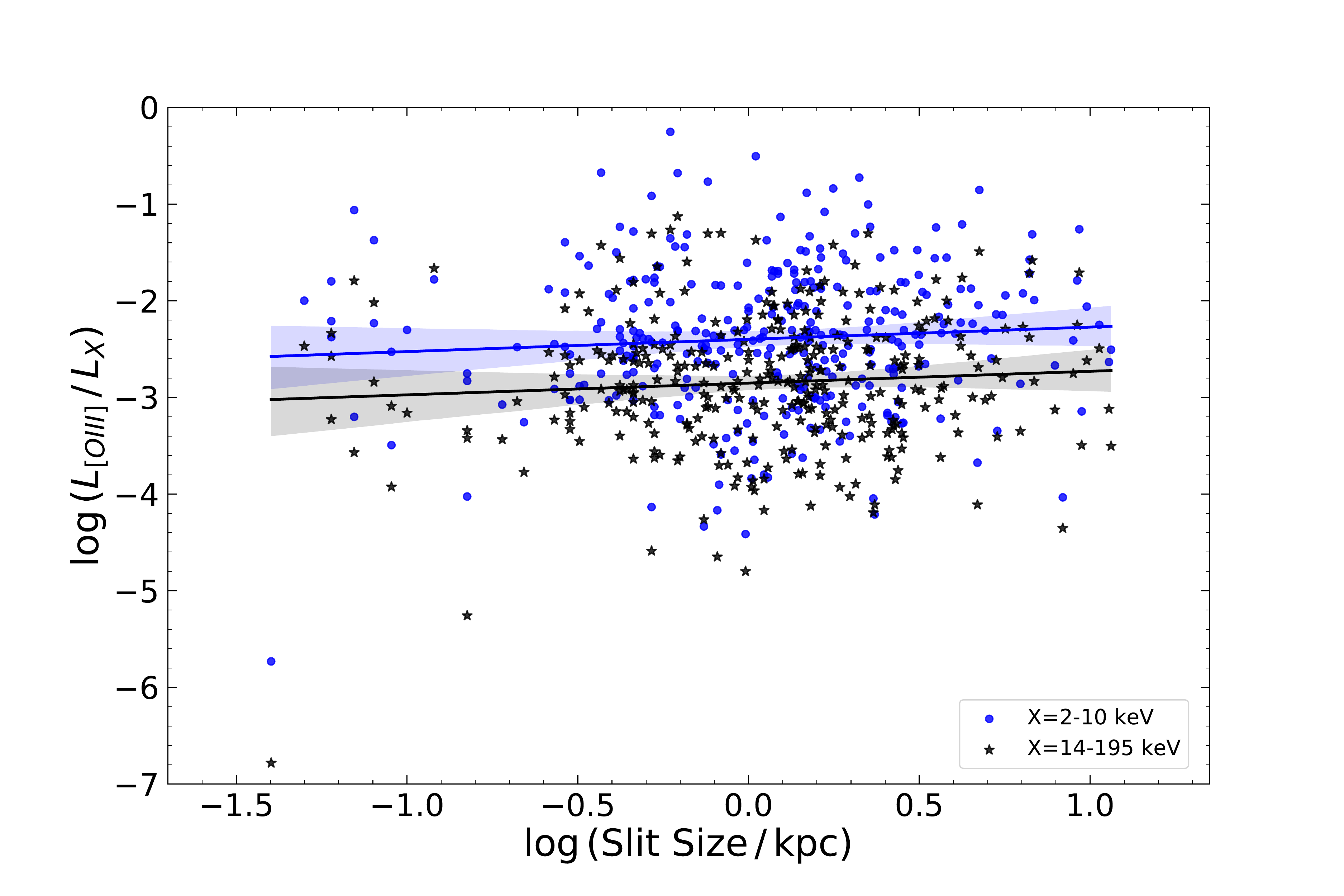}
    \caption{}
    \label{fig:OIIISS}
  \end{subfigure}
\caption{(a) The ratio of observed [O{\scriptsize\,III}] $\lambda$5007 luminosity to intrinsic X-ray luminosity vs redshift. (b) The ratio of observed [O{\scriptsize\,III}] $\lambda$5007 luminosity to intrinsic X-ray luminosity vs physical slit size (kpc) used for [O{\scriptsize\,III}] $\lambda$5007 measurements. The blue and black lines show the linear regression fit to our data for intrinsic X-ray luminosity in the 2--10 keV and 14--195 keV energy bands, respectively. The shaded regions show the uncertainty in the regression lines. The lack of correlation in both cases ($r$-value = 0.087 and 0.077, respectively) suggests that the [O{\scriptsize\,III}] $\lambda$5007 measurements are not significantly affected by the aperture size of the sources.}
\label{fig:OIII}
\end{figure*}

We have adopted observed flux measurements of the [O{\scriptsize\,III}] $\lambda$5007 emission line in our analysis. To eliminate possible aperture effects on these measurements, we investigated potential correlations between the ratio of observed [O{\scriptsize\,III}] $\lambda$5007 to intrinsic X-ray luminosity and the redshift (Figure \ref{fig:OIIIz}). We obtain a linear regression with slope = $0.14\pm0.10$ for both energy bands and $p$-value = 0.18 (0.13), for 2--10 keV (14--195 keV) X-ray luminosity. We also checked, in Figure \ref{fig:OIIISS}, if $L_{\rm [O\,III]}/L_{\rm X}$ correlates with the physical size of the slits used for the [O{\scriptsize\,III}] $\lambda$5007 measurements (reported in \citealp{2017ApJ...850...74K} and Koss et al., in prep.). We get a linear regression line with slope = $0.13\pm0.0.09$ ($0.12\pm0.09$) for 2--10 keV (14--195 keV) X-ray luminosity and $p$-value = 0.16 for both cases. Based on these results, we can conclude that no significant correlation exists between the parameters, and we can exclude aperture effects from our analysis.


\bibliographystyle{mnras}
 \bibliography{ms}

\begin{thebibliography}{117}
\expandafter\ifx\csname natexlab\endcsname\relax\def\natexlab#1{#1}\fi

\bibitem[{Ananna} et~al.(2020){Ananna}, {Treister}, {Urry}
  et~al.]{2020ApJ...889...17A}
{Ananna} T.~T., {Treister} E., {Urry} C.~M., et~al., 2020, \apj, 889, 1, 17

\bibitem[{Antonucci}(1993)]{1993ARA&A..31..473A}
{Antonucci} R., 1993, \araa, 31, 473

\bibitem[{Arnaud}(1996)]{1996ASPC..101...17A}
{Arnaud} K.~A., 1996, {XSPEC: The First Ten Years}, vol. 101 of { Astronomical
  Society of the Pacific Conference Series\/}, ~17

\bibitem[{Barthelmy} et~al.(2005){Barthelmy}, {Barbier}, {Cummings}
  et~al.]{2005SSRv..120..143B}
{Barthelmy} S.~D., {Barbier} L.~M., {Cummings} J.~R., et~al., 2005, \ssr, 120,
  3-4, 143

\bibitem[{Baumgartner} et~al.(2013){Baumgartner}, {Tueller}, {Markwardt}
  et~al.]{2013ApJS..207...19B}
{Baumgartner} W.~H., {Tueller} J., {Markwardt} C.~B., et~al., 2013, \apjs, 207,
  2, 19

\bibitem[{Beckmann} et~al.(2009){Beckmann}, {Soldi}, {Ricci}
  et~al.]{2009A&A...505..417B}
{Beckmann} V., {Soldi} S., {Ricci} C., et~al., 2009, \aap, 505, 1, 417

\bibitem[{Berney} et~al.(2015){Berney}, {Koss}, {Trakhtenbrot}
  et~al.]{2015MNRAS.454.3622B}
{Berney} S., {Koss} M., {Trakhtenbrot} B., et~al., 2015, \mnras, 454, 4, 3622

\bibitem[{Bianchi} et~al.(2010){Bianchi}, {Chiaberge}, {Evans}
  et~al.]{2010MNRAS.405..553B}
{Bianchi} S., {Chiaberge} M., {Evans} D.~A., et~al., 2010, \mnras, 405, 1, 553

\bibitem[{Bianchi} et~al.(2006){Bianchi}, {Guainazzi} \&
  {Chiaberge}]{2006A&A...448..499B}
{Bianchi} S., {Guainazzi} M., {Chiaberge} M., 2006, \aap, 448, 2, 499

\bibitem[{Bisogni} et~al.(2017){Bisogni}, {Marconi} \&
  {Risaliti}]{2017MNRAS.464..385B}
{Bisogni} S., {Marconi} A., {Risaliti} G., 2017, \mnras, 464, 1, 385

\bibitem[{Braito} et~al.(2017){Braito}, {Reeves}, {Bianchi}, {Nardini} \&
  {Piconcelli}]{2017A&A...600A.135B}
{Braito} V., {Reeves} J.~N., {Bianchi} S., {Nardini} E., {Piconcelli} E., 2017,
  \aap, 600, A135

\bibitem[{Brinkman} et~al.(2002){Brinkman}, {Kaastra}, {van der Meer}
  et~al.]{2002A&A...396..761B}
{Brinkman} A.~C., {Kaastra} J.~S., {van der Meer} R.~L.~J., et~al., 2002, \aap,
  396, 761

\bibitem[{Burrows} et~al.(2005){Burrows}, {Hill}, {Nousek}
  et~al.]{2005SSRv..120..165B}
{Burrows} D.~N., {Hill} J.~E., {Nousek} J.~A., et~al., 2005, \ssr, 120, 3-4,
  165

\bibitem[{Caglar} et~al.(2020){Caglar}, {Burtscher}, {Brandl}
  et~al.]{2020A&A...634A.114C}
{Caglar} T., {Burtscher} L., {Brandl} B., et~al., 2020, \aap, 634, A114

\bibitem[{Cappi} et~al.(2006){Cappi}, {Panessa}, {Bassani}
  et~al.]{2006A&A...446..459C}
{Cappi} M., {Panessa} F., {Bassani} L., et~al., 2006, \aap, 446, 2, 459

\bibitem[{Comastri} et~al.(1995){Comastri}, {Setti}, {Zamorani} \&
  {Hasinger}]{1995A&A...296....1C}
{Comastri} A., {Setti} G., {Zamorani} G., {Hasinger} G., 1995, \aap, 296, 1

\bibitem[{Dadina}(2007)]{2007A&A...461.1209D}
{Dadina} M., 2007, \aap, 461, 3, 1209

\bibitem[{Dadina}(2008)]{2008A&A...485..417D}
{Dadina} M., 2008, \aap, 485, 2, 417

\bibitem[{Dadina} et~al.(2010){Dadina}, {Guainazzi}, {Cappi}
  et~al.]{2010A&A...516A...9D}
{Dadina} M., {Guainazzi} M., {Cappi} M., et~al., 2010, \aap, 516, A9

\bibitem[{Eguchi} et~al.(2009){Eguchi}, {Ueda}, {Terashima}, {Mushotzky} \&
  {Tueller}]{2009ApJ...696.1657E}
{Eguchi} S., {Ueda} Y., {Terashima} Y., {Mushotzky} R., {Tueller} J., 2009,
  \apj, 696, 2, 1657

\bibitem[{Elitzur}(2012)]{2012ApJ...747L..33E}
{Elitzur} M., 2012, \apjl, 747, 2, L33

\bibitem[{Fabbiano} et~al.(2018){Fabbiano}, {Paggi}, {Karovska}, {Elvis},
  {Maksym} \& {Wang}]{2018ApJ...865...83F}
{Fabbiano} G., {Paggi} A., {Karovska} M., {Elvis} M., {Maksym} W.~P., {Wang}
  J., 2018, \apj, 865, 2, 83

\bibitem[{Fabian} et~al.(2006){Fabian}, {Celotti} \&
  {Erlund}]{2006MNRAS.373L..16F}
{Fabian} A.~C., {Celotti} A., {Erlund} M.~C., 2006, \mnras, 373, 1, L16

\bibitem[{Fabian} et~al.(2000){Fabian}, {Iwasawa}, {Reynolds} \&
  {Young}]{2000PASP..112.1145F}
{Fabian} A.~C., {Iwasawa} K., {Reynolds} C.~S., {Young} A.~J., 2000, \pasp,
  112, 775, 1145

\bibitem[{Fabian} et~al.(2008){Fabian}, {Vasudevan} \&
  {Gandhi}]{2008MNRAS.385L..43F}
{Fabian} A.~C., {Vasudevan} R.~V., {Gandhi} P., 2008, \mnras, 385, 1, L43

\bibitem[{Fabian} et~al.(2009){Fabian}, {Vasudevan}, {Mushotzky}, {Winter} \&
  {Reynolds}]{2009MNRAS.394L..89F}
{Fabian} A.~C., {Vasudevan} R.~V., {Mushotzky} R.~F., {Winter} L.~M.,
  {Reynolds} C.~S., 2009, \mnras, 394, 1, L89

\bibitem[{Feigelson} \& {Nelson}(1985)]{1985ApJ...293..192F}
{Feigelson} E.~D., {Nelson} P.~I., 1985, \apj, 293, 192

\bibitem[{Ferrarese} \& {Merritt}(2000)]{2000ApJ...539L...9F}
{Ferrarese} L., {Merritt} D., 2000, \apjl, 539, 1, L9

\bibitem[{Fukazawa} et~al.(2011){Fukazawa}, {Hiragi}, {Mizuno}
  et~al.]{2011ApJ...727...19F}
{Fukazawa} Y., {Hiragi} K., {Mizuno} M., et~al., 2011, \apj, 727, 1, 19

\bibitem[{Gebhardt} et~al.(2000){Gebhardt}, {Bender}, {Bower}
  et~al.]{2000ApJ...539L..13G}
{Gebhardt} K., {Bender} R., {Bower} G., et~al., 2000, \apjl, 539, 1, L13

\bibitem[{George} \& {Fabian}(1991)]{1991MNRAS.249..352G}
{George} I.~M., {Fabian} A.~C., 1991, \mnras, 249, 352

\bibitem[{Ghisellini} et~al.(1994){Ghisellini}, {Haardt} \&
  {Matt}]{1994MNRAS.267..743G}
{Ghisellini} G., {Haardt} F., {Matt} G., 1994, \mnras, 267, 743

\bibitem[{Gilli} et~al.(2007){Gilli}, {Comastri} \&
  {Hasinger}]{2007A&A...463...79G}
{Gilli} R., {Comastri} A., {Hasinger} G., 2007, \aap, 463, 1, 79

\bibitem[{G{\'o}mez-Guijarro} et~al.(2017){G{\'o}mez-Guijarro},
  {Gonz{\'a}lez-Mart{\'\i}n}, {Ramos Almeida}, {Rodr{\'\i}guez-Espinosa} \&
  {Gallego}]{2017MNRAS.469.2720G}
{G{\'o}mez-Guijarro} C., {Gonz{\'a}lez-Mart{\'\i}n} O., {Ramos Almeida} C.,
  {Rodr{\'\i}guez-Espinosa} J.~M., {Gallego} J., 2017, \mnras, 469, 3, 2720

\bibitem[{Greene} et~al.(2014){Greene}, {Pooley}, {Zakamska}, {Comerford} \&
  {Sun}]{2014ApJ...788...54G}
{Greene} J.~E., {Pooley} D., {Zakamska} N.~L., {Comerford} J.~M., {Sun} A.-L.,
  2014, \apj, 788, 1, 54

\bibitem[{Guainazzi} \& {Bianchi}(2007)]{2007MNRAS.374.1290G}
{Guainazzi} M., {Bianchi} S., 2007, \mnras, 374, 4, 1290

\bibitem[{Haardt} \& {Maraschi}(1991)]{1991ApJ...380L..51H}
{Haardt} F., {Maraschi} L., 1991, \apjl, 380, L51

\bibitem[{H{\"o}nig} et~al.(2014){H{\"o}nig}, {Gandhi}, {Asmus}
  et~al.]{2014MNRAS.438..647H}
{H{\"o}nig} S.~F., {Gandhi} P., {Asmus} D., et~al., 2014, \mnras, 438, 1, 647

\bibitem[{Ichikawa} et~al.(2019{\natexlab{a}}){Ichikawa}, {Kawamuro},
  {Shidatsu} et~al.]{2019ApJ...883L..13I}
{Ichikawa} K., {Kawamuro} T., {Shidatsu} M., et~al., 2019{\natexlab{a}}, \apjl,
  883, 1, L13

\bibitem[{Ichikawa} et~al.(2019{\natexlab{b}}){Ichikawa}, {Ricci}, {Ueda}
  et~al.]{2019ApJ...870...31I}
{Ichikawa} K., {Ricci} C., {Ueda} Y., et~al., 2019{\natexlab{b}}, \apj, 870, 1,
  31

\bibitem[{Ichikawa} et~al.(2012){Ichikawa}, {Ueda}, {Terashima}
  et~al.]{2012ApJ...754...45I}
{Ichikawa} K., {Ueda} Y., {Terashima} Y., et~al., 2012, \apj, 754, 1, 45

\bibitem[{Jansen} et~al.(2001){Jansen}, {Lumb}, {Altieri}
  et~al.]{2001A&A...365L...1J}
{Jansen} F., {Lumb} D., {Altieri} B., et~al., 2001, \aap, 365, L1

\bibitem[{Kawamuro} et~al.(2016){Kawamuro}, {Ueda}, {Tazaki}, {Ricci} \&
  {Terashima}]{2016ApJS..225...14K}
{Kawamuro} T., {Ueda} Y., {Tazaki} F., {Ricci} C., {Terashima} Y., 2016, \apjs,
  225, 1, 14

\bibitem[{Kewley} et~al.(2006){Kewley}, {Groves}, {Kauffmann} \&
  {Heckman}]{2006MNRAS.372..961K}
{Kewley} L.~J., {Groves} B., {Kauffmann} G., {Heckman} T., 2006, \mnras, 372,
  3, 961

\bibitem[{Kinkhabwala} et~al.(2002){Kinkhabwala}, {Sako}, {Behar}
  et~al.]{2002ApJ...575..732K}
{Kinkhabwala} A., {Sako} M., {Behar} E., et~al., 2002, \apj, 575, 2, 732

\bibitem[Kormendy \& Ho(2013)]{doi:10.1146/annurev-astro-082708-101811}
Kormendy J., Ho L.~C., 2013, Annual Review of Astronomy and Astrophysics, 51,
  1, 511

\bibitem[{Kormendy} \& {Ho}(2013)]{2013ARA&A..51..511K}
{Kormendy} J., {Ho} L.~C., 2013, \araa, 51, 1, 511

\bibitem[Kormendy \& Richstone(1995)]{doi:10.1146/annurev.aa.33.090195.003053}
Kormendy J., Richstone D., 1995, Annual Review of Astronomy and Astrophysics,
  33, 1, 581

\bibitem[{Koss} et~al.(2017){Koss}, {Trakhtenbrot}, {Ricci}
  et~al.]{2017ApJ...850...74K}
{Koss} M., {Trakhtenbrot} B., {Ricci} C., et~al., 2017, \apj, 850, 1, 74

\bibitem[{Koss} et~al.(2016){Koss}, {Assef}, {Balokovi{\'c}}
  et~al.]{2016ApJ...825...85K}
{Koss} M.~J., {Assef} R., {Balokovi{\'c}} M., et~al., 2016, \apj, 825, 2, 85

\bibitem[{Koss} et~al.(2018){Koss}, {Blecha}, {Bernhard}
  et~al.]{2018Natur.563..214K}
{Koss} M.~J., {Blecha} L., {Bernhard} P., et~al., 2018, \nat, 563, 7730, 214

\bibitem[{Koss} et~al.(2020){Koss}, {Strittmatter}, {Lamperti}
  et~al.]{2020arXiv201015849K}
{Koss} M.~J., {Strittmatter} B., {Lamperti} I., et~al., 2020, arXiv e-prints,
  arXiv:2010.15849

\bibitem[{Krimm} et~al.(2013){Krimm}, {Holland}, {Corbet}
  et~al.]{2013ApJS..209...14K}
{Krimm} H.~A., {Holland} S.~T., {Corbet} R.~H.~D., et~al., 2013, \apjs, 209, 1,
  14

\bibitem[{Krolik} et~al.(1994){Krolik}, {Madau} \&
  {Zycki}]{1994ApJ...420L..57K}
{Krolik} J.~H., {Madau} P., {Zycki} P.~T., 1994, \apjl, 420, L57

\bibitem[{Lamperti} et~al.(2017){Lamperti}, {Koss}, {Trakhtenbrot}
  et~al.]{2017MNRAS.467..540L}
{Lamperti} I., {Koss} M., {Trakhtenbrot} B., et~al., 2017, \mnras, 467, 1, 540

\bibitem[{Levenson} et~al.(2002){Levenson}, {Krolik}, {{\.Z}ycki}
  et~al.]{2002ApJ...573L..81L}
{Levenson} N.~A., {Krolik} J.~H., {{\.Z}ycki} P.~T., et~al., 2002, \apjl, 573,
  2, L81

\bibitem[{Lightman} \& {White}(1988)]{1988ApJ...335...57L}
{Lightman} A.~P., {White} T.~R., 1988, \apj, 335, 57

\bibitem[{Maddox}(2018)]{2018MNRAS.480.5203M}
{Maddox} N., 2018, \mnras, 480, 4, 5203

\bibitem[{Magdziarz} \& {Zdziarski}(1995)]{1995MNRAS.273..837M}
{Magdziarz} P., {Zdziarski} A.~A., 1995, \mnras, 273, 3, 837

\bibitem[{Magorrian} et~al.(1998){Magorrian}, {Tremaine}, {Richstone}
  et~al.]{1998AJ....115.2285M}
{Magorrian} J., {Tremaine} S., {Richstone} D., et~al., 1998, \aj, 115, 6, 2285

\bibitem[{Maiolino} et~al.(1998){Maiolino}, {Salvati}, {Bassani}
  et~al.]{1998A&A...338..781M}
{Maiolino} R., {Salvati} M., {Bassani} L., et~al., 1998, \aap, 338, 781

\bibitem[{Marchesi} et~al.(2018){Marchesi}, {Ajello}, {Marcotulli}, {Comastri},
  {Lanzuisi} \& {Vignali}]{2018ApJ...854...49M}
{Marchesi} S., {Ajello} M., {Marcotulli} L., {Comastri} A., {Lanzuisi} G.,
  {Vignali} C., 2018, \apj, 854, 1, 49

\bibitem[{Marconi} \& {Hunt}(2003)]{2003ApJ...589L..21M}
{Marconi} A., {Hunt} L.~K., 2003, \apjl, 589, 1, L21

\bibitem[{Mateos} et~al.(2016){Mateos}, {Carrera}, {Alonso-Herrero}
  et~al.]{2016ApJ...819..166M}
{Mateos} S., {Carrera} F.~J., {Alonso-Herrero} A., et~al., 2016, \apj, 819, 2,
  166

\bibitem[{Matt} et~al.(1991){Matt}, {Perola} \& {Piro}]{1991A&A...247...25M}
{Matt} G., {Perola} G.~C., {Piro} L., 1991, \aap, 247, 25

\bibitem[{Mitsuda} et~al.(2007){Mitsuda}, {Bautz}, {Inoue}
  et~al.]{2007PASJ...59S...1M}
{Mitsuda} K., {Bautz} M., {Inoue} H., et~al., 2007, \pasj, 59, S1

\bibitem[{Mushotzky} et~al.(1993){Mushotzky}, {Done} \&
  {Pounds}]{1993ARA&A..31..717M}
{Mushotzky} R.~F., {Done} C., {Pounds} K.~A., 1993, \araa, 31, 717

\bibitem[{Noguchi} et~al.(2009){Noguchi}, {Terashima} \&
  {Awaki}]{2009ApJ...705..454N}
{Noguchi} K., {Terashima} Y., {Awaki} H., 2009, \apj, 705, 1, 454

\bibitem[{Noguchi} et~al.(2010){Noguchi}, {Terashima}, {Ishino}
  et~al.]{2010ApJ...711..144N}
{Noguchi} K., {Terashima} Y., {Ishino} Y., et~al., 2010, \apj, 711, 1, 144

\bibitem[{Nucita} et~al.(2010){Nucita}, {Guainazzi}, {Longinotti},
  {Santos-Lleo}, {Maruccia} \& {Bianchi}]{2010A&A...515A..47N}
{Nucita} A.~A., {Guainazzi} M., {Longinotti} A.~L., {Santos-Lleo} M.,
  {Maruccia} Y., {Bianchi} S., 2010, \aap, 515, A47

\bibitem[{Oh} et~al.(2011){Oh}, {Sarzi}, {Schawinski} \&
  {Yi}]{2011ApJS..195...13O}
{Oh} K., {Sarzi} M., {Schawinski} K., {Yi} S.~K., 2011, \apjs, 195, 2, 13

\bibitem[{Oh} et~al.(2017){Oh}, {Schawinski}, {Koss}
  et~al.]{2017MNRAS.464.1466O}
{Oh} K., {Schawinski} K., {Koss} M., et~al., 2017, \mnras, 464, 2, 1466

\bibitem[{Oh} et~al.(2015){Oh}, {Yi}, {Schawinski}, {Koss}, {Trakhtenbrot} \&
  {Soto}]{2015ApJS..219....1O}
{Oh} K., {Yi} S.~K., {Schawinski} K., {Koss} M., {Trakhtenbrot} B., {Soto} K.,
  2015, \apjs, 219, 1, 1

\bibitem[{Page} et~al.(2005){Page}, {Reeves}, {O'Brien} \&
  {Turner}]{2005MNRAS.364..195P}
{Page} K.~L., {Reeves} J.~N., {O'Brien} P.~T., {Turner} M.~J.~L., 2005, \mnras,
  364, 1, 195

\bibitem[{Paliya} et~al.(2019){Paliya}, {Koss}, {Trakhtenbrot}
  et~al.]{2019ApJ...881..154P}
{Paliya} V.~S., {Koss} M., {Trakhtenbrot} B., et~al., 2019, \apj, 881, 2, 154

\bibitem[{Paltani} \& {Ricci}(2017)]{2017A&A...607A..31P}
{Paltani} S., {Ricci} C., 2017, \aap, 607, A31

\bibitem[{Panagiotou} \& {Walter}(2019)]{2019A&A...626A..40P}
{Panagiotou} C., {Walter} R., 2019, \aap, 626, A40

\bibitem[{Ponti} et~al.(2013){Ponti}, {Cappi}, {Costantini}
  et~al.]{2013A&A...549A..72P}
{Ponti} G., {Cappi} M., {Costantini} E., et~al., 2013, \aap, 549, A72

\bibitem[{Ramos Almeida} \& {Ricci}(2017)]{2017NatAs...1..679R}
{Ramos Almeida} C., {Ricci} C., 2017, Nature Astronomy, 1, 679

\bibitem[{Reeves} \& {Turner}(2000)]{2000MNRAS.316..234R}
{Reeves} J.~N., {Turner} M.~J.~L., 2000, \mnras, 316, 2, 234

\bibitem[{Ricci} et~al.(2018){Ricci}, {Ho}, {Fabian}
  et~al.]{2018MNRAS.480.1819R}
{Ricci} C., {Ho} L.~C., {Fabian} A.~C., et~al., 2018, \mnras, 480, 2, 1819

\bibitem[{Ricci} et~al.(2017{\natexlab{a}}){Ricci}, {Trakhtenbrot}, {Koss}
  et~al.]{2017ApJS..233...17R}
{Ricci} C., {Trakhtenbrot} B., {Koss} M.~J., et~al., 2017{\natexlab{a}}, \apjs,
  233, 2, 17

\bibitem[{Ricci} et~al.(2017{\natexlab{b}}){Ricci}, {Trakhtenbrot}, {Koss}
  et~al.]{2017Natur.549..488R}
{Ricci} C., {Trakhtenbrot} B., {Koss} M.~J., et~al., 2017{\natexlab{b}}, \nat,
  549, 7673, 488

\bibitem[Ricci et~al.(2015)Ricci, Ueda, Koss, Trakhtenbrot, Bauer \&
  Gandhi]{Ricci_2015}
Ricci C., Ueda Y., Koss M.~J., Trakhtenbrot B., Bauer F.~E., Gandhi P., 2015,
  The Astrophysical Journal, 815, 1, L13

\bibitem[{Ricci} et~al.(2014){Ricci}, {Ueda}, {Paltani}, {Ichikawa}, {Gand hi}
  \& {Awaki}]{2014MNRAS.441.3622R}
{Ricci} C., {Ueda} Y., {Paltani} S., {Ichikawa} K., {Gand hi} P., {Awaki} H.,
  2014, \mnras, 441, 4, 3622

\bibitem[{Ricci} et~al.(2011){Ricci}, {Walter}, {Courvoisier} \&
  {Paltani}]{2011A&A...532A.102R}
{Ricci} C., {Walter} R., {Courvoisier} T.~J.~L., {Paltani} S., 2011, \aap, 532,
  A102

\bibitem[{Risaliti} et~al.(1999){Risaliti}, {Maiolino} \&
  {Salvati}]{1999ApJ...522..157R}
{Risaliti} G., {Maiolino} R., {Salvati} M., 1999, \apj, 522, 1, 157

\bibitem[{Risaliti} et~al.(2011){Risaliti}, {Salvati} \&
  {Marconi}]{2011MNRAS.411.2223R}
{Risaliti} G., {Salvati} M., {Marconi} A., 2011, \mnras, 411, 4, 2223

\bibitem[{Rojas} et~al.(2020){Rojas}, {Sani}, {Gavignaud}
  et~al.]{2020MNRAS.491.5867R}
{Rojas} A.~F., {Sani} E., {Gavignaud} I., et~al., 2020, \mnras, 491, 4, 5867

\bibitem[{Rybicki} \& {Lightman}(1986)]{1986rpa..book.....R}
{Rybicki} G.~B., {Lightman} A.~P., 1986, {Radiative Processes in Astrophysics}

\bibitem[{Sako} et~al.(2000){Sako}, {Kahn}, {Paerels} \&
  {Liedahl}]{2000ApJ...543L.115S}
{Sako} M., {Kahn} S.~M., {Paerels} F., {Liedahl} D.~A., 2000, \apjl, 543, 2,
  L115

\bibitem[{Sarzi} et~al.(2006){Sarzi}, {Falc{\'o}n-Barroso}, {Davies}
  et~al.]{2006MNRAS.366.1151S}
{Sarzi} M., {Falc{\'o}n-Barroso} J., {Davies} R.~L., et~al., 2006, \mnras, 366,
  4, 1151

\bibitem[{Schawinski} et~al.(2015){Schawinski}, {Koss}, {Berney} \&
  {Sartori}]{2015MNRAS.451.2517S}
{Schawinski} K., {Koss} M., {Berney} S., {Sartori} L.~F., 2015, \mnras, 451, 3,
  2517

\bibitem[{Schurch} et~al.(2004){Schurch}, {Warwick}, {Griffiths} \&
  {Kahn}]{2004MNRAS.350....1S}
{Schurch} N.~J., {Warwick} R.~S., {Griffiths} R.~E., {Kahn} S.~M., 2004,
  \mnras, 350, 1, 1

\bibitem[{Shimizu} et~al.(2017){Shimizu}, {Mushotzky}, {Mel{\'e}ndez}, {Koss},
  {Barger} \& {Cowie}]{2017MNRAS.466.3161S}
{Shimizu} T.~T., {Mushotzky} R.~F., {Mel{\'e}ndez} M., {Koss} M.~J., {Barger}
  A.~J., {Cowie} L.~L., 2017, \mnras, 466, 3, 3161

\bibitem[{Shu} et~al.(2010){Shu}, {Yaqoob} \& {Wang}]{2010ApJS..187..581S}
{Shu} X.~W., {Yaqoob} T., {Wang} J.~X., 2010, \apjs, 187, 2, 581

\bibitem[{Smith} et~al.(2020){Smith}, {Mushotzky}, {Koss}
  et~al.]{2020MNRAS.492.4216S}
{Smith} K.~L., {Mushotzky} R.~F., {Koss} M., et~al., 2020, \mnras, 492, 3, 4216

\bibitem[{Tanaka} et~al.(1994){Tanaka}, {Inoue} \& {Holt}]{1994PASJ...46L..37T}
{Tanaka} Y., {Inoue} H., {Holt} S.~S., 1994, \pasj, 46, L37

\bibitem[{Tanimoto} et~al.(2018){Tanimoto}, {Ueda}, {Kawamuro}, {Ricci},
  {Awaki} \& {Terashima}]{2018ApJ...853..146T}
{Tanimoto} A., {Ueda} Y., {Kawamuro} T., {Ricci} C., {Awaki} H., {Terashima}
  Y., 2018, \apj, 853, 2, 146

\bibitem[{Tremaine} et~al.(2002){Tremaine}, {Gebhardt}, {Bender}
  et~al.]{2002ApJ...574..740T}
{Tremaine} S., {Gebhardt} K., {Bender} R., et~al., 2002, \apj, 574, 2, 740

\bibitem[{Turner} et~al.(1997){Turner}, {George}, {Nandra} \&
  {Mushotzky}]{1997ApJS..113...23T}
{Turner} T.~J., {George} I.~M., {Nandra} K., {Mushotzky} R.~F., 1997, \apjs,
  113, 1, 23

\bibitem[{Ueda} et~al.(2014){Ueda}, {Akiyama}, {Hasinger}, {Miyaji} \&
  {Watson}]{2014ApJ...786..104U}
{Ueda} Y., {Akiyama} M., {Hasinger} G., {Miyaji} T., {Watson} M.~G., 2014,
  \apj, 786, 2, 104

\bibitem[{Ueda} et~al.(2003){Ueda}, {Akiyama}, {Ohta} \&
  {Miyaji}]{2003ApJ...598..886U}
{Ueda} Y., {Akiyama} M., {Ohta} K., {Miyaji} T., 2003, \apj, 598, 2, 886

\bibitem[{Ueda} et~al.(2007){Ueda}, {Eguchi}, {Terashima}
  et~al.]{2007ApJ...664L..79U}
{Ueda} Y., {Eguchi} S., {Terashima} Y., et~al., 2007, \apjl, 664, 2, L79

\bibitem[{Ueda} et~al.(2015){Ueda}, {Hashimoto}, {Ichikawa}
  et~al.]{2015ApJ...815....1U}
{Ueda} Y., {Hashimoto} Y., {Ichikawa} K., et~al., 2015, \apj, 815, 1, 1

\bibitem[{Urry} \& {Padovani}(1995)]{1995PASP..107..803U}
{Urry} C.~M., {Padovani} P., 1995, \pasp, 107, 803

\bibitem[{Vasudevan} \& {Fabian}(2009)]{2009MNRAS.392.1124V}
{Vasudevan} R.~V., {Fabian} A.~C., 2009, \mnras, 392, 3, 1124

\bibitem[{Vasudevan} et~al.(2009){Vasudevan}, {Mushotzky}, {Winter} \&
  {Fabian}]{2009MNRAS.399.1553V}
{Vasudevan} R.~V., {Mushotzky} R.~F., {Winter} L.~M., {Fabian} A.~C., 2009,
  \mnras, 399, 3, 1553

\bibitem[{Veilleux} \& {Osterbrock}(1987)]{1987ApJS...63..295V}
{Veilleux} S., {Osterbrock} D.~E., 1987, \apjs, 63, 295

\bibitem[{Vietri} et~al.(2018){Vietri}, {Piconcelli}, {Bischetti}
  et~al.]{2018A&A...617A..81V}
{Vietri} G., {Piconcelli} E., {Bischetti} M., et~al., 2018, \aap, 617, A81

\bibitem[{Vignali} et~al.(1999){Vignali}, {Comastri}, {Cappi}, {Palumbo},
  {Matsuoka} \& {Kubo}]{1999ApJ...516..582V}
{Vignali} C., {Comastri} A., {Cappi} M., {Palumbo} G.~G.~C., {Matsuoka} M.,
  {Kubo} H., 1999, \apj, 516, 2, 582

\bibitem[{Weisskopf} et~al.(2000){Weisskopf}, {Tananbaum}, {Van Speybroeck} \&
  {O'Dell}]{2000SPIE.4012....2W}
{Weisskopf} M.~C., {Tananbaum} H.~D., {Van Speybroeck} L.~P., {O'Dell} S.~L.,
  2000, {Chandra X-ray Observatory (CXO): overview}, vol. 4012 of { Society of
  Photo-Optical Instrumentation Engineers (SPIE) Conference Series\/},  2--16

\bibitem[{Winter} et~al.(2009){Winter}, {Mushotzky}, {Reynolds} \&
  {Tueller}]{2009ApJ...690.1322W}
{Winter} L.~M., {Mushotzky} R.~F., {Reynolds} C.~S., {Tueller} J., 2009, \apj,
  690, 2, 1322

\bibitem[{Yamada} et~al.(2020){Yamada}, {Ueda}, {Tanimoto}
  et~al.]{2020ApJ...897..107Y}
{Yamada} S., {Ueda} Y., {Tanimoto} A., et~al., 2020, \apj, 897, 1, 107

\bibitem[{Yaqoob} \& {Padmanabhan}(2004)]{2004ApJ...604...63Y}
{Yaqoob} T., {Padmanabhan} U., 2004, \apj, 604, 1, 63

\bibitem[{Young} et~al.(2001){Young}, {Wilson} \&
  {Shopbell}]{2001ApJ...556....6Y}
{Young} A.~J., {Wilson} A.~S., {Shopbell} P.~L., 2001, \apj, 556, 1, 6

\bibitem[{Zappacosta} et~al.(2018){Zappacosta}, {Comastri}, {Civano}
  et~al.]{2018ApJ...854...33Z}
{Zappacosta} L., {Comastri} A., {Civano} F., et~al., 2018, \apj, 854, 1, 33

\end{thebibliography}

\section*{Affiliations}

$^{1}$N\'ucleo de Astronom\'ia de la Facultad de Ingenier\'ia, Universidad Diego Portales, Av. Ej\'ercito Libertador 441, Santiago 22, Chile\\
$^{2}$Kavli Institute for Astronomy and Astrophysics, Peking University, Beijing 100871, People's Republic of China\\
$^{3}$George Mason University, Department of Physics \& Astronomy, MS 3F3, 4400 University Drive, Fairfax, VA 22030, USA\\
$^{4}$Department of Astronomy, Kyoto University, Kitashirakawa-Oiwake-cho, Sakyo-ku, Kyoto 606-8502, Japan\\
$^{5}$National Astronomical Observatory of Japan, 2-21-1 Osawa, Mitaka, Tokyo 181-8588, Japan\\
$^{6}$Eureka Scientific, 2452 Delmer Street Suite 100, Oakland, CA 94602-3017, USA\\
$^{7}$School of Physics and Astronomy, Tel Aviv University, Tel Aviv 69978, Israel\\
$^{8}$Korea Astronomy \& Space Science institute, 776, Daedeokdae-ro, Yuseong-gu, Daejeon 34055, Republic of Korea\\
$^{9}$Instituto de Astrofísica and Centro de Astroingeniería, Facultad de Física, Pontificia Universidad Católica de Chile, Casilla 306,\\ Santiago 22, Chile\\
$^{10}$Millennium Institute of Astrophysics, Nuncio Monseñor Sótero Sanz 100, Providencia, Santiago, Chile\\
$^{11}$Space Science Institute, 4750 Walnut Street, Suite 205, Boulder, Colorado 80301, USA\\
$^{12}$Department of Astronomy, University of Florida, 211 Bryant Space Science Center, Gainesville, FL 32611, USA\\
$^{13}$National Radio Astronomy Observatory, 520 Edgemont Rd, Charlottesville, VA 22903, USA\\
$^{14}$Osservatorio Astronomico di Roma, via di Frascati 33, I-00078 Monte Porzio Catone, Italy\\
$^{15}$Jet Propulsion Laboratory, California Institute of Technology, 4800 Oak Grove Drive, MS 169-224, Pasadena, CA 91109, USA\\
$^{16}$European Southern Observatory, Alonso de Cordova 3107, Casilla 19, Santiago 19001, Chile\\
$^{17}$Department of Astronomy, University of Maryland, College Park, MD 20742, USA\\
$^{18}$Leiden Observatory, PO Box 9513, 2300 RA, Leiden, The Netherlands\\
$^{19}$Frontier Research Institute for Interdisciplinary Sciences, Tohoku University, Sendai 980-8578, Japan\\
$^{20}$International Centre for Radio Astronomy Research, M468, The University of Western Australia, 35 Stirling Highway, Crawley, \\
WA 6009, Australia\\
$^{21}$Australia Telescope National Facility, CSIRO Astronomy and Space Science, P.O. Box 76, Epping, NSW 1710, Australia\\
$^{22}$Institute of Particle Astrophysics and Cosmology, Stanford University, 452 Lomita Mall, Stanford, CA 94305, USA \\
$^{23}$Yale Center for Astronomy \& Astrophysics, Physics Department, New Haven, CT 06520, USA\\
$^{24}$Cahill Center for Astronomy and Astrophysics, California Institute of Technology, Pasadena, CA 91125, USA\\

\label{lastpage}

 \end{document}